\documentclass[
    aps,
    prb,
    reprint,
    superscriptaddress,
    nofootinbib,
 longbibliography
]{revtex4-2}

\usepackage{amsmath,amssymb,amsfonts}
\usepackage{bm}
\usepackage{graphicx}
\usepackage{physics}
\usepackage{braket}
\usepackage{bbm}
\usepackage[dvipsnames]{xcolor}

\usepackage[
  colorlinks=true,
  linkcolor=magenta,   
  citecolor=magenta,   
  urlcolor=magenta     
]{hyperref}
\usepackage[normalem]{ulem}
\usepackage{xcolor}

\let\dotlessi\i
\renewcommand{\i}{%
  \ifmmode
    \mathrm{i}%
  \else
    \dotlessi
  \fi
}

\renewcommand{\Tr}{\operatorname{Tr}}

\newcommand{\id}{\mathbbm{1}}

\begin{document}

\title{
Local time-reversal-invariant topological markers in two dimensions
}

\author{Thomas Klein Kvorning}
\affiliation{Department of Physics, KTH Royal Institute of Technology, 106 91, Stockholm, Sweden}
\author{Miguel F.\ Martínez}
\affiliation{Department of Physics, KTH Royal Institute of Technology, 106 91, Stockholm, Sweden}
\author{Julia D.\ Hannukainen}
\affiliation{T.C.M. Group, Cavendish Laboratory, J.J. Thomson Avenue, Cambridge CB3 0US, United Kingdom}

\date{\today}

\begin{abstract}
Local topological markers provide a real-space characterization of topological phases and are particularly suited to inhomogeneous and noncrystalline models.
We derive a local topological marker, the \textit{Chern-Simons descendant marker}, for characterizing the topology of two-dimensional time-reversal-invariant near-Gaussian states directly from the one-particle density matrix \(\rho\). 
We construct an auxiliary three-dimensional state characterized by the local Chern-Simons marker, and use dimensional reduction to obtain the Chern-Simons descendant marker for the original two-dimensional state.
This derivation introduces a local Hermitian involution \(S\) that is odd under time reversal.
When \([S,\rho]=0\), the local  Chern-Simons descendant marker reduces to half the difference of the local Chern markers in the \(S=\pm1\) sectors.
Constructing an \(S\) that commutes exactly with \(\rho\) generally requires fine-tuning, but this is not needed---the local Chern-Simon marker remains well defined whenever \(\Gamma=\{\rho,S\}-S\) has a spectral gap around zero, even when \([S,\rho]\neq0\).
This leaves the practical problem of identifying a suitable \(S\) for a given state, which, importantly, is straightforward even when spin is not a candidate. 
We show this by considering both strong Rashba and strong Dresselhaus spin-orbit coupling, for which rudimentary knowledge of the physical origin of the state reduces the general problem of choosing $S$ to a one-parameter family \(S(\theta)\).
We evaluate the optimal angle $\theta$ maximizing the spectral gap of $\Gamma$ analytically assuming translation invariance, and find that the gap of $\Gamma$ remains open for a broad range of angles around the optimum.
The same, analytic, value of \(\theta\) remains generally valid in disordered systems, providing an easy way to obtain choice of \(S\) for evaluating the local Chern-Simons descendant marker of the state.
Our results demonstrate that suitable choices of \(S\) are readily available for physically relevant models and require no fine tuning, making the local Chern-Simons descendant  marker a practical tool for characterizing the topology of states directly in real space.

\end{abstract}

\maketitle

\section{Introduction}

Given a Gaussian state, what is its topological phase?
Hamiltonian-based classifications determine the topological phase indirectly by identifying the relevant symmetries and evaluating the appropriate topological invariant for the states below a spectral gap~\cite{ryu2010,schnyder08,ludwig15,hasan2012,qi2011,moessner_moore_2021}.
But within a fixed symmetry class, the topology is a property of the Gaussian state itself~\cite{chen2010}.
The one-particle density matrix~\cite{penrose1956,bera2015,bera2017} of a Gaussian state completely specifies the occupied single-particle subspace that encodes the topological information.
The topology of a state can therefore be characterized without referring to the spectrum of a parent Hamiltonian.
This motivates the use of topological invariants formulated directly in terms of the one-particle density matrix.

Local topological markers is a term describing a diverse collection of real-space topological invariants~\cite{prodan2010,loring2010,bianco2011,prodan2011,loring2015,huang2018,hughes2019,loring2019,hofstetter2019,kleinkvorning2020,loring2020, schulz-baldes2021,markov2021, barnett2021,cerjan2022,
hannukainen2022,jezequel2022,munoz2022, chen2023, hannukainen2024,cerjan2024,jezequel2024,jezequel2025,martinez2026, jezequel2026}, particularly suited to systems without translation symmetry, including disordered, amorphous, and other noncrystalline systems~\cite{groth2009,kraus2012,bandres2016,mansha2017,agarwala2017,poyhonen2018, mitchell2018, brzezinska2018, yanbin2019, varjas2019, grushin2020,ivaki2020,marsal2020,sahlberg2020,agarwala2020,mukati2020,focassio2021,hughes2021, skipetrov2021,spring2021,wang2022,corbae2023,schirmann2025, gomezpaz2026leveraging, ghosh2026}.
Among these markers, the local Chern, chiral, and Chern-Simons markers form a category of markers which are formulated directly in terms of the one-particle density matrix and therefore characterize the topology of the state itself~\cite{bianco2011, hannukainen2022, hannukainen2024}.
The local Chern marker is a \(\mathbb Z\) valued invariant that characterizes integer Chern phases in even spatial dimensions~\cite{kitaev2006, prodan2010, bianco2011, hofstetter2019, dOrnellas2022}.
In odd spatial dimensions, the local chiral and Chern–Simons markers characterize all phases classified by ($\mathbb Z$) and ($\mathbb Z_2$), respectively~\cite{hannukainen2022, hannukainen2024}.

An important feature of the local Chern, chiral, and Chern-Simons markers is that they are not only local markers, they are also spatially resolved markers, meaning that they assign a topological marker throughout real space, rather than only a single bulk invariant~\cite{bianco2011}.
The point wise marker need not itself be quantized, but coarse graining over a finite region can determine whether that region supports a well-defined local topological phase~\cite{bianco2011}.
Varying the size of the coarse-graining region therefore reveals both spatially inhomogeneous topological regions and the length scale on which the marker approaches a quantized value~\cite{wang2024,salib2025}.
In this sense, space-resolved topological markers probe not only the topological phase, but also its spatial organization and the scale on which a local topological description becomes meaningful.

For two-dimensional time-reversal-invariant states, time-reversal symmetry forces the total Chern number of the occupied subspace to vanish, even in a nontrivial quantum spin Hall phase~\cite{kane2005, kane2005b, fu2006, sheng2006}.
There are several real-space formulations for the \(\mathbb Z_2\) invariant characterizing such states, which differ both in the information they require as input, and in the form of locality they provide~\cite{prodan2009, loring2015,akagi2017, huang2018, huang2018_2, li2019,loring2019,doll2021, ailardoni2022, chen2023, bau2024, gomezpaz2026leveraging}.
The aim of this work is to construct a local topological marker that both complements existing real-space methods, and offers additional capabilities: A practical to use, state-based formulation that is spatially resolved throughout real space, and characterizes Gaussian and near-Gaussian states directly from the one-particle density matrix.

In this work we introduce a state-based and spatially resolved local topological \(\mathbb Z_2\) marker in even dimensions directly from the one-particle density matrix.
In two dimensions, we derive this local marker by dimensional reduction of the three-dimensional local Chern-Simons marker~\cite{hannukainen2022}.
The resulting topological marker is the local Chern-Simons descendant marker---the local CS-descendant marker for short.
The dimensional reduction introduces a local Hermitian operator \(S\) satisfying \(S^2=1\), and is odd under time-reversal symmetry.
If \(S\) commutes with the one-particle density matrix, $\rho$, the occupied subspace decomposes into the \(S=+1\) and \(S=-1\) eigenspaces, which are exchanged by time reversal.
The local Chern marker is evaluated separately in the two subspaces, and their difference, modulo two, determines the local CS-descendant marker.
If  \(S\) is chosen to be a conserved physical spin operator, the two subspaces are simply the spin sectors and the resulting expression coincides with the local spin-Chern marker~\cite{prodan2009,bau2024}.
But the dimensional-reduction derivation of the local CS-descendant marker does not identify \(S\) with spin: \(S\) can be any local Hermitian involution that is odd under time reversal.
Although the commutation relation $[S,\rho]=0$ simplifies the evaluation, it is not a requirement for the CS-descendant to be well defined~\cite{prodan2009,bau2024}.
We show that the marker remains well defined for \([S,\rho]\neq0\), provided that the corresponding auxilliary operator,  \(\Gamma=\{\rho,S\}-S\) is spectrally gapped around zero.

The freedom in the choice of \(S\) raises a practical question: how should \(S\) be identified for a given state?
We provide a systematic procedure for determining \(S\) in physically relevant settings.
In many physical settings the origin of the state identifies a small family of local operators compatible with its spin-orbital structure.
The optimal choice of \(S\) within this family of possible operators is the one that maximizes the spectral gap of \(\Gamma\).

We illustrate how \(S\) is identified and optimized in a sequence of time-reversal-invariant models with increasing spin mixing.
We explain that an operator acting jointly on the spin and orbital degrees of freedom may provide an exactly conserved \(S\), even when physical spin itself is not conserved.
Local spin-frame disorder~\cite{schirmann2025, gomezpaz2026leveraging}, corresponding to spatially dependent rotations of the spin quantization axis, generate spatially varying operators \(S(\mathbf r)\).
These local operators are inserted directly into the local CS-descendant marker, adapting it to states whose spin-orbital structure varies across the system.
For models in which Rashba and Dresselhaus couplings~\cite{bychkov1984, manchon2015, winkler2003, zutic2004, kane2005, prodan2009} break physical spin conservation, we construct a one-parameter family of candidate operators \(S(\theta)\) from the spin-orbital structure.
For a translation-invariant system, maximizing the spectral gap of \(\Gamma\) over \(\theta\) determines the optimal angle analytically, while a broad range of nearby angles retains a sizable gap.
Without translation invariance, the same parameter family \(S(\theta)\) provides a real-space ansatz whose optimal angle follows from maximizing the auxiliary spectral gap directly in real space.
We obtain the optimal value of $S$ in models with structural and scalar on site potential disorder, where translation symmetry is absent,  and use it to evaluate the local CS-descendant marker.
For completeness, we explain how a fully variational search over local time-reversal-odd operators is possible in principle when no simple physically motivated family can be identified, although the low-dimensional approach is considerably more practical.
We show how the variational approach becomes trivial to implement in the presence of strong local disorder.

The results presented in this work place two-dimensional time-reversal-invariant \(\mathbb Z_2\) phases within the same state-based local-marker framework as the Chern, chiral, and Chern-Simons markers, while removing the need to identify \(S\) with a conserved spin operator.
The resulting formulation remains practical in spin-mixed and noncrystalline systems through physically motivated and real-space-optimized choices of \(S\).
Beyond identifying the \(\mathbb Z_2\) phase, the CS-descendant marker provides direct access to how topology is spatially organized and to the length scales on which a local topological description is meaningful.

The remainder of the paper is organized as follows.
Section~II introduces the one-particle density matrix formulation of local topology and reviews the local Chern marker.
Section~III derives the local CS-descendant marker; readers interested primarily in the final marker formula may proceed directly to Sec.~III\,C.
Section~IV develops physically motivated choices of the local operator \(S\) entering the expression for the CS-descendant marker in the presence of increasingly general forms of spin mixing.
Section~VI concludes with a discussion of the resulting framework and its extensions.

\section{Local topology from the one-particle density matrix}
A fermionic Gaussian state $|\Psi\rangle$ is completely specified by its two-point correlation functions.
The normal correlations \(\tilde{\rho}_{ij}=\langle\Psi|c_i^\dagger c_j|\Psi\rangle\), and anomalous correlations \(\kappa_{ij}=\langle\Psi|c_i c_j|\Psi\rangle\), where $c_i^\dagger$, and $c_j$ are fermionic creation and annhiliation operators, combine into the Bogoliubov-de Gennes one-particle density matrix~\cite{penrose1956,bera2015,bera2017,lezama2017,kells2018}.
\begin{align}
    \varrho
    &=
    \begin{pmatrix}
       \tilde{\rho} & \kappa \\
        \kappa^\dagger & 1-\tilde{\rho} ^*
    \end{pmatrix}.
    \label{eq:full-opdm}
\end{align}
For a Gaussian state, \(\varrho\) is a projector, $\varrho^2= \varrho$, with eigenvalues zero or one.
The corresponding eigenspaces play the role of the unoccupied and occupied single-particle subspaces which specify the Gaussian state.
The matrix elements of $\varrho$ of local Gaussian states decay exponentially with the distance between the corresponding sites, 
\begin{align}
|\varrho_{ij}|\sim \text{exp}\left(-\frac{|\mathbf r_i-\mathbf r_j|}{\xi_\varrho}\right),
\label{eq:local-rho}
\end{align}
 where \(\xi_\varrho\) is the single-particle correlation length.

For a local Gaussian \(U(1)\) symmetry, the corresponding single-particle symmetry is represented by a local Hermitian involution \(\tau_z\), that commutes with the one-particle density matrix.
For particle number conservation, \(\tau_z\) represents the particle-hole Pauli $z$-matrix in the Bogoliubov-de Gennes basis of Eq.~\eqref{eq:full-opdm}. 
The commutation relation $[\tau_z,\rho]=0$ forces \(\kappa=0\), so that  \(\varrho\) separates into particle and hole blocks.
Within each symmetry reduced block, time-reversal, particle-hole, and chiral constraints determine the corresponding Altland-Zirnbauer class~\cite{cartan26,zirnbauer96,altland97} and hence the allowed topological classification~\cite{kitaev09}.
For $U(1)$ charge conservation the two blocks are related according to the diagonal blocks in Eq.~\eqref{eq:local-rho}, which means that both blocks have the same topology.
For general $U(1)$ symmetries this need not be the case; the blocks can have independent $\mathbb{Z}_2$ and $\mathbb{Z}$ invariant and different blocks for the same state can belong to different Altland-Zirnbauer symmetry classes.
To simplify notation we denote the symmetry reduced block entering the topological classification by $\rho$.
Thus, \(\rho=\varrho\) when no unitary symmetry is resolved, while for a particle-number-conserving state \(\rho=\tilde{\rho}\) in the particle block.

For a translation-invariant state,  $\rho$ is resolved in crystal momentum $\mathbf k$ as \(\rho(\mathbf k)\).
At each \(\mathbf k\), its image defines the occupied single-particle subspace, and these subspaces vary continuously across the Brillouin zone to form a vector bundle~\cite{kitaev09,nakahara18}.
Topological invariants characterize the global structure of this family of occupied subspaces together with the relevant symmetry constraints~\cite{ryu2010,schnyder08,ludwig15}.
For example, a nonzero Chern number obstructs a globally smooth and periodic choice of basis for the occupied subspace.

The local Chern, chiral, and Chern-Simons markers provide real-space counterparts of momentum-space topological invariants in even and odd spatial dimensions~\cite{bianco2011, hannukainen2022,hannukainen2024}.
In this work we focus on topology in even dimensions; in particular in two dimensions, where the integer topological phases in Altland-Zirnbauer classes A, D, and C are characterized by a Chern number~\cite{ryu2010}.
The corresponding real-space invariant is the local Chern marker~\cite{bianco2011},
\begin{align}
    \mathcal C(\mathbf r)
    &=
    2\pi\i
    \sum_\alpha
    \varepsilon^{i_1i_2}
    \left[
        \rho X_{i_1}
        \rho X_{i_2}
        \rho
    \right]_{(\mathbf r,\alpha),(\mathbf r,\alpha)},
    \label{eq:local-chern-marker}
\end{align}
where \(\varepsilon^{i_1i_2}\) is the antisymmetric Levi-Civita symbol, \(i_1,i_2\in\{x,y\}\) are summed over, \(X_i\) denotes the diagonal position operators, and \(\alpha\) labels the local internal degrees of freedom.
The local character of Eq.~\eqref{eq:local-chern-marker} relies on the locality of the one-particle density matrix \(\rho\),\eqref{eq:local-rho}.

In a translation-invariant bulk, summing over the internal degrees of freedom makes \(\mathcal C(\mathbf r)\) independent of the unit cell, and its value reproduces the Chern number~\cite{bianco2011}.
When translation symmetry is absent, crystal momentum and the Brillouin-zone vector bundle are no longer available, but the local Chern marker in Eq.~\eqref{eq:local-chern-marker} remains well defined, provided that $\rho$ is exponentially localized, Eq.~(\ref{eq:local-rho}).
The value of the local marker depends on location and varies from point to point, resolving the spatial contributions to the topological invariant.
Its pointwise value need not be quantized; instead, the spatial profile resolves local contributions to the topological invariant.
Coarse graining over progressively larger bulk regions suppresses short-distance fluctuations and reveals the scale on which the local marker approaches a quantized value.
The size of the region required for convergence defines a length scale \(\xi_{\mathrm{top}}\) on which the local topological phase becomes well defined.
Spatial variations of the local marker therefore distinguish a homogeneous topological phase with a large \(\xi_{\mathrm{top}}\) from interfaces, defects, or finite domains with a different local topology.

The one-particle-density-matrix formulation of local topological markers admits an extension beyond Gaussian states~\cite{hannukainen2024}.
For an interacting state, the eigenvalues of the one-particle density matrix, its natural orbitals, need not be restricted to zero and one~\cite{bera2015,bera2017}.
If this occupation spectrum remains gapped, it still separates predominantly occupied and unoccupied natural orbitals.
This separation defines the flattened projector
\begin{align}
    \rho_{\mathrm f}
    &=
    \frac{1}{2}
    \left[
        1+
        |2\rho-1|^{-1}(2\rho-1)
    \right],
\end{align}
which leaves the natural orbitals unchanged, and satisfies \(\rho_{\mathrm f}^2=\rho_{\mathrm f}\).
Because \(\rho_{\mathrm f}\) is a spectral function of \(\rho\), it inherits its symmetry constraints.
For a local one-particle density matrix with a finite occupation gap, spectral flattening preserves locality and therefore allows the local Chern, chiral, and Chern-Simons markers to be evaluated from the flattened $\rho_{\mathrm f}$, thereby characterizing the topology of the underlying state when it is near Gaussian~\cite{hannukainen2024}.

\section{The local Chern-Simons descendant marker}

In two dimensions, five Altland-Zirnbauer classes support nontrivial topological phases~\cite{ryu2010}.
The local Chern marker characterizes the integer-valued phases in classes A, D, and C and is given by Eq.~\eqref{eq:local-chern-marker}.
The remaining two classes are characterized by a \(\mathbb Z_2\) topological invariant: Class AII is constrained by time-reversal symmetry, while class DIII additionally has a particle-hole constraint and, consequently, a chiral constraint.
In this section, we derive the local CS-descendant marker for class AII using dimensional reduction of the local Chern-Simons marker in three dimensions~\cite{qi2008,ryu2010}.
The derivation for class DIII is analogous, but with the symmetry-resolved projector given by the generalized one-particle density matrix \(\varrho\), Eq~\eqref{eq:full-opdm}.

The first two subsections derive the local CS-descendant marker, starting from dimensional reduction and subsequently extending the result to a general form.
The final subsection summarizes the resulting expressions and the conditions under which they apply.

\subsection{Dimensional reduction of the local Chern-Simons marker}

We consider a two-dimensional Gaussian state with one-particle density matrix \(\rho\).
The density matrix obeys a time-reversal constraint $T\rho^*T^\dagger=  \rho$, where \(\mathcal T=T\mathcal K\) with \(\mathcal T^2=-1\),  is the time-reversal operator in real space.
Here the unitary operator \(T=\i\sigma_y\) where $\sigma_y$ is a Pauli matrix, and \(\mathcal K\) denotes complex conjugation.
The state therefore belongs to class AII in the Altland-Zirnbauer classification, and is characterized by a $\mathbb{Z}_2$ invariant.

To derive the local CS-descendant marker for two-dimensional class-AII states, we start from a three-dimensional class-CII state with the same \(\mathbb Z_2\) topology, for which the local marker is known.
States in class AII in two dimensions are related to states in class CII in three dimensions through a Bott extension~\cite{ryu2010}.
The homotopy classes of two-dimensional class-AII states are in one-to-one correspondence with those of three-dimensional class-CII states under Bott periodicity.
This means that, although AII and CII are distinct symmetry classes, the correspondence preserves their \(\mathbb Z_2\) topology: trivial states map to trivial states and nontrivial states to nontrivial states.
A Bott extension provides a topology preserving map between the two-dimensional state in class AII and a state in class CII  by introducing an auxiliary dimension \(k_z\).
The $\mathbb{Z}_2$ invariant for class CII in three dimensions is known, it is the local Chern-Simons marker~\cite{hannukainen2022}.
The topology of the original two-dimensional state is therefore characterized by evaluating the three-dimensional local Chern-Simons marker for the one particle density matrix of the Bott-extended three-dimensional state.
Since the associated local Chern-Simons marker is expressed in terms of the auxiliary dimension through \(k_z\), we perform a dimensional reduction by integrating out the auxiliary degrees of freedom.
The resulting two-dimensional expression defines the local CS-descendant marker of the original two-dimensional AII state, written entirely in terms of its one-particle density.

A convenient realization of the Bott extension~\cite{chiu2016} maps the two-dimensional class-AII one-particle density matrix \(\rho\) to
\begin{align}
    \widetilde\rho(k_z)
    &=
    \frac{1}{2}
    \left[
        \id
        -
        \sin k_z\,(\id-2\rho)\otimes\tau_x
        -
        \cos k_z\,\id\otimes\tau_z
    \right],
    \label{eq:bott-extension}
\end{align}
where \(k_z\) parametrizes the auxiliary dimension and \(\tau_a\), \(a=x,y,z\), are Pauli matrices acting in an auxiliary two-dimensional space.
The extended one-particle density matrix in Eq.~\eqref{eq:bott-extension} inherits the time-reversal constraint of the two-dimensional state.
The antiunitary operator
\begin{align}
    \widetilde{\mathcal T}
    &=
    (T\otimes\tau_z)\mathcal K
\end{align}
implements time reversal in the extended space and satisfies
\begin{align}
    \widetilde{\mathcal T}
    \widetilde\rho(-k_z)
    \widetilde{\mathcal T}^{-1}
    &=
    \widetilde\rho(k_z).
\end{align}
The extended state also satisfies the particle-hole constraint
\begin{align}
    \widetilde{\mathcal C}
    &=
    (\id\otimes\tau_y)\widetilde{\mathcal T},
\end{align}
which exchanges the occupied and unoccupied subspaces,
\begin{align}
    \widetilde{\mathcal C}
    \widetilde\rho(-k_z)
    \widetilde{\mathcal C}^{-1}
    &=
    \id-\widetilde\rho(k_z).
\end{align}
Together, the time-reversal and particle-hole constraints give the
chiral operator
\begin{align}
    \widetilde S_{\rm CII}
    &=
    \widetilde{\mathcal C}\widetilde{\mathcal T}
    =
    -\id\otimes\tau_y ,
    \label{eq:chiral_operator_CII}
\end{align}
which satisfies
\begin{align}
    \{\widetilde S_{\rm CII},\widetilde \rho (k_z)\}=\widetilde S_{\rm CII}
\end{align}
These constraints place the extended state in class CII.

Since \(\rho^2=\rho\), the extended one-particle density matrix in Eq.~\eqref{eq:bott-extension} also satisfies $\widetilde\rho(k_z)^2=\widetilde\rho(k_z)$ for every \(k_z\).
\(\widetilde\rho(k_z)\) inherits the spatial locality of \(\rho\), since the extension only combines \(\rho\) with local operators acting in the auxiliary space.
The Bott extension therefore defines a localized Gaussian one-particle density matrix throughout the auxiliary dimension.

The topology of the auxilliary state in class CII is characterized by the local Chern-Simons marker~\cite{hannukainen2022}
\begin{align}
    \nu_{\rm CS}(\mathbf r)
    &=
    \frac{8\pi\i}{3}\!
    \int\!\!\frac{dk_z}{2\pi}\!
    \sum_{\beta}\!\varepsilon^{ijk}\!
    \left[
        \widetilde\rho\widetilde S X_i
        \widetilde\rho X_j
        \widetilde\rho X_k
        \widetilde\rho
\right]_{\substack{
    \smash{\raisebox{0.2ex}{\(\textstyle \bmod 2,\)}}\\
    \smash{\raisebox{-1.1ex}{\hspace{-0.9em}
    \(\scriptstyle (\mathbf r,\beta),(\mathbf r,\beta)\)}}
}}
    \label{eq:bott-chern-simons-marker}
\end{align}
The Latin indices $i=1,2,3$ are summed over, where \(X_1\) and \(X_2\) are the physical position operators, \(X_3=\i\partial_{k_z}\) acts along the auxiliary dimension.
Here \(\beta=(\alpha,\lambda)\) combines the local internal index \(\alpha\) defining internal degrees of freedom of the original two-dimensional state with the auxiliary index \(\lambda=\pm 1\), see Eq.~(\ref{eq:bott-extension}), which are traced over.
The operator \(\widetilde S\) in Eq.~\eqref{eq:bott-chern-simons-marker} is any local operator satisfying
\begin{subequations}\label{eq:Stilde-conditions}
\begin{align}
    &\widetilde S^\dagger=\widetilde S,
    \label{eq:Stilde-hermitian}
    \\
    &\widetilde S^2=1,
    \label{eq:Stilde-squared}
    \\
    &\widetilde{\mathcal T}\widetilde S
    \widetilde{\mathcal T}^{-1}
    =-\widetilde S ,
    \label{eq:Stilde-trs}
    \\
    &\{\widetilde S,\widetilde \rho\}=\widetilde S .
    \label{eq:Stilde-anticommutator}
\end{align}
\end{subequations}
The canonical chiral constraint of class CII, Eq.~\eqref{eq:chiral_operator_CII}, is even under time reversal and therefore cannot serve as \(\widetilde S\) in Eq.~\eqref{eq:bott-chern-simons-marker}.
A convenient time-reversal-odd candidate for \(\widetilde S\) is obtained from the canonical CII chiral constraint by introducing a local Hermitian operator \(S\) acting in the original two-dimensional space.
Up to an overall sign, this gives
\begin{align}
    \widetilde S
    &=
    S\otimes\tau_y .
    \label{eq:Stilde-S}
\end{align}
The conditions in Eq.~\eqref{eq:Stilde-conditions} determine the corresponding requirements on \(S\).
Eqs.~\eqref{eq:Stilde-hermitian} and \eqref{eq:Stilde-squared} require
\begin{align}
    S^\dagger
    &=
    S,
    &
    S^2
    &=
    1 .
\end{align}
Under time reversal, $ \widetilde{\mathcal T}\widetilde S
    \widetilde{\mathcal T}^{-1}=
    \left(
        T S^* T^\dagger
    \right)\otimes\tau_y $,
so Eq.~\eqref{eq:Stilde-trs} requires \(S\) to also be odd under time reversal:
\begin{align}
    TS^*T^\dagger
    &=
    -S .
    \label{eq:S-time-reversal}
\end{align}
The chiral condition in Eq.~\eqref{eq:Stilde-anticommutator} determines the remaining requirement on $S$.
Using Eqs.~\eqref{eq:bott-extension} and \eqref{eq:Stilde-S},
\begin{align}
    \{\widetilde S,\widetilde\rho(k_z)\}
    &=
    \widetilde S
    +
    \sin k_z\,[S,\rho]\otimes\tau_y\tau_x ,
    \label{eq:Stilde-rho-anticommutator_1}
\end{align}
\begin{align}
    \{\widetilde S,\widetilde\rho(k_z)\}
    =
    \widetilde S
    \quad\Longleftrightarrow\quad
    [S,\rho]=0 .
    \label{eq:Stilde-rho-anticommutator2}
\end{align}
The time-reversal condition in Eq.~\eqref{eq:S-time-reversal}, together with \(S^2=1\), pairs the \(S=+1\) and \(S=-1\) eigenspaces and consequently implies
\(\Tr S=0\).

To obtain a local marker in the original two-dimensional space, we carry out a dimensional reduction by first tracing over the auxiliary space labeled by $\lambda$, and then integrating the auxiliary dimension $k_z$. 
Each factor of \(\widetilde\rho\), Eq.~\eqref{eq:bott-extension}, in the Chern-Simons marker, Eq.~\eqref{eq:bott-chern-simons-marker},  contains an identity term together with terms proportional to \(\tau_x\) and \(\tau_z\) in the auxiliary space.
The antisymmetrization in the Chern-Simons marker eliminates every contribution in which at least one of the four factors of \(\widetilde\rho\) contributes with its identity term.
Therefore, within the antisymmetrized product of the Chern-Simons marker, each factor of \(\widetilde\rho\) contributes with either a \(\tau_x\) or a \(\tau_z\) matrix in the auxiliary space.
Together with the contribution $\tau_y$ from the operator $\tilde S$, the trace over the auxiliary space of Eq.~(\ref{eq:bott-chern-simons-marker}) becomes
\begin{align}
    \sum_{\lambda=\pm}
    \left[
        \tau_y
        \tau_{a_1}\tau_{a_2}\tau_{a_3}\tau_{a_4}
    \right]_{(\lambda)(\lambda)},
    \qquad
    a_n\in\{x,z\}.
    \label{eq:Trace-tau}
\end{align}
The Pauli-matrix algebra makes the sum over \(\lambda\) nonzero only when $ \tau_{a_1}\tau_{a_2}\tau_{a_3}\tau_{a_4} \propto\tau_y .$
This implies that only terms containing an odd number of \(\tau_x\) factors and an odd number of \(\tau_z\) factors survive the sum over the auxiliary index.

After tracing over the auxiliary space, the Chern-Simons marker in Eq.~\eqref{eq:bott-chern-simons-marker} takes the intermediate form
\begin{align}
    \nu_{\rm CS}(\mathbf r)
    &=
    8\pi\i
    \int \frac{dk_z}{2\pi}\,
    f(k_z)
    \sum_\alpha
    \varepsilon^{ij}
    \left[
        \rho S X_i
        \rho S X_j
        \rho S
    \right]_{\substack{
        \smash{\raisebox{0.2ex}{\(\textstyle \bmod 2,\)}}\\
        \smash{\raisebox{-1.1ex}{\hspace{-0.9em}
        \(\scriptstyle (\mathbf r,\alpha),(\mathbf r,\alpha)\)}}
    }}
    \label{eq:intermediate-bott-marker}
\end{align}
where the Latin indices \(i=1,2\), and $f(k_z)=\sin^2 k_z$.
The details of the calculation of the trace are given in Appendix~\ref{app:dimensional-reduction}.
Performing the  integral over the auxiliary dimension gives the local topological CS-descendant marker in two dimensions:
\begin{align}
    \nu_{\rm CSD}(\mathbf r)
    &=
    \pi\i\,\mathcal 
    \sum_\alpha
    \varepsilon^{ij}
    \left[
        \rho S X_i
        \rho S X_j
        \rho S
\right]_{\substack{
    \smash{\raisebox{0.2ex}{\(\textstyle \bmod 2,\)}}\\
    \smash{\raisebox{-0ex}{\hspace{0em}
    \(\scriptstyle (\mathbf r,\alpha),(\mathbf r,\alpha)\)}}
}}
    \label{eq:conserved-S-marker-general}
\end{align}

The local CS-descendant marker in Eq.~\eqref{eq:conserved-S-marker-general} can be written as the difference of two Chern markers, since the condition \(S^2=1\) splits
the local Hilbert space into two eigensectors of \(S\) with eigenvalues \(\pm1\).
The corresponding projectors
\begin{align}
    P_\pm
    &=
    \frac{1\pm S}{2}
\end{align}
resolve the one-particle density matrix into the two sectors
\(\rho_\pm=P_\pm\rho P_\pm\).
Using the commutation relation \([S,\rho]=0\) gives
\begin{align}
    \rho
    &=
    \rho_+ + \rho_-,
    &
    \rho S
    &=
    \rho_+ - \rho_- ,
\end{align}
and since  \(S\) acts locally in the internal space, the projectors \(P_\pm\) commute with the position operators, such that
\begin{align}
    \rho S X_i
    \rho S X_j
    \rho S
    &=
    \rho_+X_i\rho_+X_j\rho_+
    -
    \rho_-X_i\rho_-X_j\rho_- .
    \label{eq:sector-reduction-main}
\end{align}
The local CS-descendant marker  Eq.~\eqref{eq:conserved-S-marker-general} therefore assumes the form of the local Chern marker, eq.~\eqref{eq:local-chern-marker} in each eigenspace of \(S\),
\begin{align}
    \mathcal C_\pm(\mathbf r)
    &=
    2\pi\i
    \sum_\alpha
    \varepsilon^{ij}
    \left[
        \rho_\pm X_i
        \rho_\pm X_j
        \rho_\pm
    \right]_{(\mathbf r,\alpha),(\mathbf r,\alpha)},
    \label{eq:sector-chern-markers-main}
\end{align}
such that
\begin{align}
    \nu_{\rm CSD}(\mathbf r)
    &=
    \frac{1}{2}
    \left[
        \mathcal C_+(\mathbf r)
        -
        \mathcal C_-(\mathbf r)
    \right]
    \mod 2 .
    \label{eq:marker-sector-form-main}
\end{align}

The choice \(\widetilde S=S\otimes\tau_y\), together with the condition \(\{\widetilde S,\widetilde \rho\}=\rho\), requires \(S\) to commute with \(\rho\).
The two eigenspaces of \(S\) therefore define independent sectors of the occupied subspace, and the two-dimensional CS-descendant marker can be expressed as the difference of the corresponding Chern markers modulo two.
A conserved physical spin operator provides a special case of Eq.~\eqref{eq:marker-sector-form-main}, in which these eigenspaces coincide with the spin-up and spin-down sectors and the CS-descendant marker reduces to the spin-Chern number. 

\subsection{Defining the Chern-simons descendant marker when $[S,\rho]\neq0$ }

The Chern-Simons marker in Eq.~\eqref{eq:bott-chern-simons-marker} requires the chiral operator \(\widetilde S\) in the extended three dimensional space to obey $ \{\widetilde S,\widetilde \rho \}=\widetilde S$, which is equivalent to the condition $[S,\rho]=0$ for a state in two dimensions, Eq.~(\ref{eq:Stilde-rho-anticommutator2}).
A generic time-reversal-invariant state, however, need not admit a local time-reversal-odd involution \(S\) that commutes with \(\rho\).
To define the local CS-descendant marker for $[S,\rho]\neq0$, we seek another one-particle density matrix \(\rho'\) that is in the same topological phase as \(\rho\) and satisfies
\begin{align}
    [S,\rho']
    &=
    0 .
\end{align}
Since $\rho$, and $\rho'$ are in the same topological phase, they are necessarily connected by a continuous path of exponentially local time-reversal-invariant one-particle density matrices, so that the topology does not change along the path.
To construct such a path, we introduce the auxiliary Hermitian operator
\begin{align}
    N_t
    &=
    1
    -
    2\left[
        (1-t)\rho
        +
        tS\rho S
    \right],
    \qquad
    0\leq t\leq\frac12,
    \label{eq:interpolation-Nt}
\end{align}
which is exponentially local, and preserves time-reversal symmetry, throughout the interpolation. 
To obtain a path of one-particle density matrices from \(N_t\), we define
\begin{align}
    \rho_t
    &=
    \frac{1}{2}
    \left(
        1-\frac{N_t}{|N_t|}
    \right);
    \label{eq:rho-t}
\end{align}
when \(N_t\) has no zero eigenvalues, or equivalently when \(N_t^2>0\), the normalization \(N_t/|N_t|\) is well defined and obeys \((N_t/|N_t|)^2=1\).
This implies that \(\rho_t^2=\rho_t\), so \(\rho_t\) defines the one-particle density matrix of a Gaussian state, provided $N_t^2>0$.
Using  \(\rho^2=\rho\) and \(S^2=1\) yields
\begin{align}
    N_t^2
    &=
    (1-2t)^2
    +
    4t(1-t)\Gamma^2,
    \label{eq:interpolation-gap}
\end{align}
where
\begin{align}
    \Gamma
    &=
    \{\rho,S\}-S .
    \label{eq:Gamma}
\end{align}
For \(0\leq t<1/2\), the first term in Eq.~\eqref{eq:interpolation-gap} is strictly positive, so \(N_t\) cannot have a zero eigenvalue.
Since $ N_{1/2}^2= \Gamma^2 $, the path $\rho_t$, Eq.~\eqref{eq:rho-t}, remains well defined as long as \(\Gamma\) has a spectral gap around zero.

The two endpoints of the interpolation $\rho_t$ define the one-particle density matrices $\rho_0=\rho$,  and 
\begin{equation}
\rho´\equiv \rho_{1/2}=
    \frac{1}{2}
    \left(
        1-\frac{\Gamma}{|\Gamma|}
    \right)\quad \Gamma^2>0,
    \label{eq:rh-prime}
\end{equation}
 which belong to the same topological phase.
Since \(N_{1/2}\) commutes with \(S\), so does any function of \(N_{1/2}\), fulfilling the condition $[S,\rho']=0$.
The topology of the original state with density matrix $\rho$ is therefore characterized by the local CS-descendant marker for \(\rho'\),
\begin{align}
    \nu_{\rm CSD}(\mathbf r)
    &=
    \pi\i
    \sum_\alpha
    \varepsilon^{ij}
    \left[
        \rho' S X_i
        \rho' S X_j
        \rho' S
    \right]_{(\mathbf r,\alpha),(\mathbf r,\alpha)}
    \mod 2 .
    \label{eq:rho-prime-marker}
\end{align}

Substituting Eq.~\eqref{eq:rh-prime} into the CS-descendant marker Eq.~\eqref{eq:rho-prime-marker} produces three factors of \(S+\Gamma/|\Gamma|\) and an overall factor \(1/8\).
Simplifying the resulting expression by using \(S^2=1\), \([S,X_i]=0\), and \([S,\Gamma]=0\),  results in the expression
\begin{align}
    \nu_{\rm CSD}(\mathbf r)
    &=
    \frac{\pi\i}{8}
    \sum_\alpha
    \varepsilon^{ij}
    \left[
        \frac{\Gamma}{|\Gamma|}
        X_i
        \frac{\Gamma}{|\Gamma|}
        X_j
        \frac{\Gamma}{|\Gamma|}
    \right]_{(\mathbf r,\alpha),(\mathbf r,\alpha)}
    \mod 2 ,
    \label{eq:general-local-marker}
\end{align}
which is well defined for any local time reversal invariant involution $S$ for which $\Gamma$, Wq.~\eqref{eq:Gamma} has a gapped spectrum.

\subsection{Summary of the local CS-descendant marker}

For a general time-reversal-invariant state in two dimensions characterized by the one-particle density matrix $\rho$, the local CS-descendant marker is determined by
\begin{align}
    \Gamma
    &=
    \{\rho,S\}-S ,
\end{align}
where \(S\) is a local Hermitian involution that is odd under time reversal.
The CS-descendant marker is well defined provided that \(\Gamma\) remains spectrally gapped around zero, and takes the form
\begin{align}
    \nu_{\rm CSD}(\mathbf r)
    &=
    \frac{\pi\i}{8}
    \sum_\alpha
    \varepsilon^{ij}
    \left[
        \frac{\Gamma}{|\Gamma|}
        X_i
        \frac{\Gamma}{|\Gamma|}
        X_j
        \frac{\Gamma}{|\Gamma|}
    \right]_{(\mathbf r,\alpha),(\mathbf r,\alpha)}
    \mod 2 .
    \label{eq:general-local-marker-summary}
\end{align}
When \(S\) commutes with the one-particle density matrix, \([S,\rho]=0\), the general expression reduces to
\begin{align}
    \nu_{\rm CSD}(\mathbf r)
    &=
    \pi\i
    \sum_\alpha
    \varepsilon^{ij}
    \left[
        \rho S X_i
        \rho S X_j
        \rho S
    \right]_{(\mathbf r,\alpha),(\mathbf r,\alpha)}
    \mod 2 .
    \label{eq:commuting-local-marker-summary}
\end{align}
In this commuting limit, the two eigenspaces of \(S\) define independent occupied sectors, and the marker equals one half of the difference between their local Chern markers.

\section{The operator \(S\) in the presence of spin-orbit coupling}

The local marker in Eq.~(\ref{eq:general-local-marker}) requires a local Hermitian time-reversal-odd involution \(S\) for which \(\Gamma=\{\rho,S\}-S\), Eq.~(\ref{eq:Gamma}), remains spectrally gapped around zero.
The task is to identify such an operator for a given state.
Although $S$ may be associated with the physical spin operator in certain cases, spin conservation is not a requirement to yield a well-defined CS-descendant marker.
In this section we develop physically motivated choices of \(S\), starting from exact spin-orbital symmetries and subsequently introducing increasingly general forms of spin mixing.

\subsection{Beyond physical spin}

We consider a class-AII topological insulator with two orbitals per site and a spin-1/2 degree of freedom, defined on an arbitrary two-dimensional arrangement of sites~\cite{bernevig2006, agarwala2017, roy2026}.
The corresponding second-quantized Hamiltonian takes the form~\citep{gomezpaz2026leveraging}
\begin{equation}
     H = \sum_{jk} \mathbf{c}^\dagger_j   T_{jk} \mathbf{c}_k 
  + \sum_j  \epsilon_j  \mathbf{c}^\dagger_j \mathbf{c}_k ,
  \label{eq:model}
\end{equation}
\begin{equation}
    T_{jk}= e^{-(\delta_{jk} - r_0)/r_0} \, \Theta(R - \delta_{jk}) t_{jk},
    \label{eq:hopping-general}
\end{equation}
\begin{equation}
     \epsilon_j = M \tau_z \otimes s_0 + W_j \tau_0 \otimes s_0,
        \label{eq:onsite}
\end{equation}
where the Pauli matrices \(\tau_i\) and \(s_i\), \(i=x,y,z\), act on the orbital and spin degrees of freedom, respectively, and \(\Theta(x)\) denotes the Heaviside step function.
The fermionic creation and annihilation operators at site \(j\) are collected into the four-component spinors $\mathbf{c}^\dagger_j=(c^\dagger_{ja \uparrow}, c^\dagger_{jb \uparrow}, c^\dagger_{ja \downarrow}, c^\dagger_{jb \downarrow})^T$ and $\mathbf{c}_j=(c_{ja \uparrow}, c_{jb \uparrow}, c_{ja \downarrow}, c_{jb \downarrow})^T$ respectively, where \(a,b\) label the orbital degrees of freedom and \(\uparrow,\downarrow\) the spin degrees of freedom.
The first term in Eq.~\eqref{eq:model} controls the hopping between neighboring sites located at positions $\mathbf{r}_j$ and $\mathbf{r}_k$, related by a vector
\begin{align}
\boldsymbol{\delta}_{jk}= \mathbf{r}_k - \mathbf{r}_j \equiv \delta_{jk}  ( \hat{\mathbf x}\cos \varphi_{jk} + \hat{\mathbf y} \sin \varphi_{jk})
\end{align}
Eq.~(\ref{eq:hopping-general}) gives the explicit real-space form of the hopping matrix $T_{jk}$, which depends on $r_0$, the typical distance between sites, and $R$, a cutoff setting the maximum hopping distance~\cite{agarwala2017}.
The internal hopping matrix $t_{jk}$ introduces an orbital (O) and a spin-orbit coupling term (SO),
\begin{equation}
    \label{eq:hopping}
    t_{jk} = t_{jk}^{\rm O} + t_{jk}^{\rm SO}= \frac{A}{2} \left( \tau_z \otimes s_0 - i \tau_x \otimes (\hat{\boldsymbol{\delta}}_{jk} \vdot \mathbf{s} ) \right),
\end{equation}
where $A$ is a parameter of the model and $\hat{\boldsymbol{\delta}}_{jk} = \boldsymbol{\delta_{jk} }/ \delta_{jk}$.
The second term in Eq.~(\ref{eq:model}) is an onsite term, Eq.~(\ref{eq:onsite}), with $M$ an onsite energy and $W_j$ a random onsite potential drawn from the uniform distribution $[-W/2, W/2]$ at each site.

The Hamiltonian in Eq.~\eqref{eq:model} conserves particle number,  so the ground state one-particle density matrix \(\rho\) corresponds to the block $\tilde{\rho}$ of the  Bogoliubov-de Gennes one-particle density matrix in Eq.~\eqref{eq:full-opdm}.
$\rho$ has a time-reversal symmetry $\rho = \mathcal{T}^\dagger \rho^* \mathcal{T}$ with $\mathcal{T}=is_y$.  
For generic \(W\neq0\), no additional antiunitary symmetry is present, and the state belongs to class AII.
At \(W=0\), the Hamiltonian, Eq.~\eqref{eq:model}, has and additional particle-hole constraint \(\mathcal C=C\mathcal K\), with \(C=\tau_y\otimes s_y\), which imposes $C\rho^*C^\dagger =1-\rho $.
Since \(\mathcal T^2=-1\) and \(\mathcal C^2=+1\), the ground state belongs to class DIII at \(W=0\)---the random on-site potential breaks particle-hole symmetry reducing the symmetry class from DIII to AII for generic \(W\neq0\)~\cite{gomezpaz2026leveraging}.
In the translation-invariant limit of a square lattice, the half-filled state described by \(\rho\) is topological for \(0<|M|<4\) and \(W=0\)~\cite{bernevig2006}.

The spin-orbit coupling term in Eq.~\eqref{eq:hopping} breaks conservation of the physical spin, $  [s_z,H] \neq 0$.
The local spin-orbital operator
\begin{align}
    S_0 = \tau_z\otimes s_z
    \label{eq:S_0}
\end{align}
is nevertheless conserved, $[S_0,H] =0$, which implies that  \([S_0,\rho]=0\).
\(S_0\) is a local Hermitian involution that is odd under time reversal, so its two eigenspaces provide the time-reversed decomposition of the occupied subspace used in
Eq.~\eqref{eq:sector-chern-markers-main}.
The topology of the state is therefore characterized by the local CS-descendant marker in Eq.~\eqref{eq:conserved-S-marker-general},
with \(S=S_0\).
The origin of the conserved operator \(S_0\) follows from the more general local operator
\begin{align}
    S(\hat{\mathbf n})
    &=
    \tau_z\otimes(\hat{\mathbf n}\cdot\mathbf s) .
\end{align}
The commutator between $S(\hat{\mathbf n})$ and the spin-orbit term is
\begin{align}
    \left[
        \tau_z \otimes (\hat{\mathbf n}\cdot\mathbf s),
        \tau_x \otimes
        (\hat{\boldsymbol{\delta}}_{jk}\cdot\mathbf s)
    \right]
    &=
    2(\hat{\mathbf n}\cdot\hat{\boldsymbol{\delta}}_{jk})
    \tau_z\tau_x .
\end{align}
Since \(\hat{\boldsymbol{\delta}}_{jk}\) lies in the \(xy\)-plane, choosing \(\hat{\mathbf n}=\hat{\mathbf z}\) makes this commutator vanish, recovering the operator \(S_0=\tau_z\otimes s_z\).
 
\subsection{Spin-frame disorder}

The operator \(S_0\) that commutes with the one-particle density matrix of Eq.~\eqref{eq:model} assumes a globally fixed spin frame.
In structurally disordered and amorphous geometries, where the notion of a globally preferred internal-space direction is lost, the local spin frame may instead vary in space~\cite{schirmann2025, gomezpaz2026leveraging}.
Such spatial variation defines spin-frame disorder, represented by local unitary rotations \(U_{\mathbf r}\) acting on the internal degrees of freedom at each site,
\begin{align}
    U &= \bigoplus_{\mathbf r} U_{\mathbf r}.
\end{align}
This results in a modified ground state one-particle density matrix and time-reversal symmetry,
\begin{align}
    \rho' &= U\rho U^\dagger,
    &
     \mathcal{T}' &= U\mathcal{T} U^\dagger.
\end{align}
The local rotations therefore change not only the state, but also the representation of the time reversal constraint.
If the reference time-reversal operator \(\mathcal T=\i s_y\) is known, the transformed operator \(\mathcal T'\) identifies the local spin frame generated by \(U\).
The same transformation is applied to the conserved operator \(S_0\),
\begin{align}
    S_0' &= U S_0 U^\dagger .
\end{align}
Because \(U\) acts locally, \(S_0'\) remains a local Hermitian involution, is odd under the transformed time-reversal symmetry, and satisfies
\begin{align}
    [S_0',\rho']
    &=
    U[S_0,\rho]U^\dagger
    =
    0 .
\end{align}
The topology of \(\rho'\) is therefore characterized by the local \(\mathbb Z_2\) marker, Eq.~\eqref{eq:marker-sector-form-main}, with \(S=S_0'\).
The operators \(S_0\) and \(S_0'\) show that the local \(\mathbb Z_2\) marker, Eq.~\eqref{eq:conserved-S-marker-general}, does not require \(S\) to coincide with a conserved physical spin or with a globally fixed internal-space direction.

\subsection{Rashba spin-orbit coupling}

Rashba spin--orbit coupling arises when inversion symmetry is extrinsically broken, for example at an interface or heterojunction~\cite{bychkov1984, zutic2004, manchon2015, winkler2003}.
In contrast to a local rotation of the spin frame, Rashba coupling introduces a new physical term into the Hamiltonian.
It therefore provides a genuine form of spin mixing and breaks the exact conservation of \(S_0=\tau_z\otimes s_z\), Eq.~\eqref{eq:S_0}.
Nevertheless, the structure of the Rashba coupling does not require a general search for \(S\).
Instead, the Rashba coupling singles out one additional local operator, which together with \(S_0\) spans a two-dimensional operator space.
The choice of \(S\) therefore reduces to a one-parameter family \(S(\theta)\), where \(\theta\) controls the relative weight of these two operator directions.
The optimal choice of \(S\) corresponds to the angle $\theta$ for which the the spectral gap of \(\Gamma=\Gamma(\theta)\) is maximal. 

The Rashba contribution to the internal hopping matrix is
\begin{align}
    t_{jk}^{\rm R}
    &=
    -\i\lambda_{\rm R}
    \hat{\boldsymbol\eta}_{jk}\cdot\mathbf s,
    \label{eq:rashba}
\end{align}
with
\begin{align}
\hat{\boldsymbol\eta}_{jk}
=
\hat{\mathbf x}\sin\varphi_{jk}
-
\hat{\mathbf y}\cos\varphi_{jk},
\label{eq:rashba-direction}
\end{align}
which breaks the conservation of \(S_0\), since
\begin{align}
    [S_0,t_{jk}^{\rm R}]
    &=
    2\lambda_{\rm R}\tau_z
    \left(
        \hat{\boldsymbol\delta}_{jk}\cdot\mathbf s
    \right)
    \neq 0 .
    \label{eq:commutator-rashba}
\end{align}
At finite Rashba coupling, we therefore introduce an additional local operator, \(\tau_y\otimes s_0\), where the two operators $S_0$, and \(\tau_y\otimes s_0\), have complementary commutation relations with the spin-dependent hopping terms.
The operator \(S_0\) commutes with the original spin-orbit hopping but not with the Rashba hopping, whereas \(\tau_y\otimes s_0\) commutes with the Rashba hopping but not with the original spin-orbit hopping:
\begin{align}
    [S_0,t_{jk}^{\rm SO}]&=0,
    &
    [\tau_y\otimes s_0,t_{jk}^{\rm R}]&=0 .
\end{align}
Moreover, the two nonzero commutators have the same internal matrix structure, but with opposite signs,
\begin{align}
    [\tau_y\otimes s_0,t_{jk}^{\rm SO}]
    &=
    -\frac{A}{\lambda_{\rm R}}
    [S_0,t_{jk}^{\rm R}] .
\end{align}
Since \(S_0\) and \(\tau_y\otimes s_0\) are Hermitian, time-reversal-odd anticommuting involutions, we define the one parameter family 
\begin{align}
    S(\theta)
    &=
    S_0\cos\theta
    +
    \tau_y\otimes s_0\sin\theta,
    \label{eq:s_theta}
\end{align}
where the spin-orbit coupling,  $t_{jk}^{\rm SO}$ favors \(\theta=0\), for which \(S=S_0\), whereas the Rashba coupling $t_{jk}^{\rm R}$ favors \(\theta=\pi/2\), for which
\(S=\tau_y\otimes s_0\).
The combined spin-dependent hopping favors the angle
\begin{align}
    \theta_{\rm s}=\arctan\left(\frac{\lambda_{\rm R}}{A}\right),
\end{align}
for which $[S(\theta_{\rm s}),t_{jk}^{\rm SO}+t_{jk}^{\rm R}]=0 $.
The full Hamiltonian, Eq.~\eqref{eq:model}, however, also contains spin-independent terms proportional to \(\tau_z\), which only commute with \(S(\theta)\) for \(\theta=0\).
The spin-dependent and spin-independent parts of the Hamiltonian, Eq.~\eqref{eq:model}, therefore favor different values of \(\theta\), so no single angle exactly conserves \(S(\theta)\) for finite Rashba coupling.
But since the CS-decendant marker does not require exact conservation of \(S\), the relevant criterion is to choose the angle that maximizes the spectral
gap of \(\Gamma(\theta)\).

The one-parameter family \(S(\theta)\) in Eq.~\eqref{eq:s_theta} is defined entirely in terms of local onsite operators and does not rely on translation symmetry.
We only invoke translation invariance to analytically determine the angle \(\theta\) that maximizes the spectral gap of \(\Gamma(\theta)\),and subsequently use this angle as an ansatz for disordered systems.
In the translation-invariant limit, the one-particle density matrix decomposes into independent momentum blocks,
\begin{align}
    \rho
    &=
    \bigoplus_{\mathbf k}\rho(\mathbf k).
\end{align}
Since \(S(\theta)\) acts only on the internal degrees of freedom, \(\Gamma(\theta)\) inherits the same decomposition,
\begin{align}
    \Gamma(\theta)
    &=
    \bigoplus_{\mathbf k}
    \Gamma(\mathbf k,\theta).
\end{align}
Determining the angle that maximizes the spectral gap of \(\Gamma(\theta)\) therefore reduces to analyzing the spectrum of \(\Gamma(\mathbf k,\theta)\) at each momentum \(\mathbf k\).
The Bloch Hamiltonian corresponding to Eq.~\eqref{eq:model}, including the Rashba term in Eq.~\eqref{eq:rashba}, is
\begin{align}
    H(\mathbf k)
    &=
    d(\mathbf k)\tau_z
    +
    \tau_x\,\mathbf g(\mathbf k)\cdot\mathbf s
    +
    \mathbf h_{\rm R}(\mathbf k)\cdot\mathbf s ,
    \label{eq:block-ham-rashba}
\end{align}
where
\begin{align}
    \mathbf g(\mathbf k)
    &=
    2A
    \begin{pmatrix}
        \sin k_x\\
        \sin k_y
    \end{pmatrix},
    &
    \mathbf h_{\rm R}(\mathbf k)
    &=
    2\lambda_{\rm R}
    \begin{pmatrix}
        \sin k_y\\
        -\sin k_x
    \end{pmatrix}.
\end{align}
The two spin-dependent vectors are orthogonal at every momentum and have a momentum-independent relative magnitude,
\begin{align}
    \mathbf g(\mathbf k)\cdot\mathbf h_{\rm R}(\mathbf k)
    &=0,
    &
    \frac{|\mathbf h_{\rm R}(\mathbf k)|}
         {|\mathbf g(\mathbf k)|}
    &=
    \frac{\lambda_{\rm R}}{A}.
    \label{eq:rasha-hg-orthogonal}
\end{align}
The orthogonality implies that the corresponding spin operators anticommute, $ \{
        \hat{\mathbf g}(\mathbf k)\cdot\mathbf s,
        \hat{\mathbf h}_{\rm R}(\mathbf k)\cdot\mathbf s
    \}=0$, which together with the anticommutation relation \(\{\tau_z,\tau_x\}=0\), this makes
\begin{align}
    K_{\rm R}(\mathbf k)
    &=
    \tau_z\,
    \hat{\mathbf h}_{\rm R}(\mathbf k)\cdot\mathbf s
\end{align}
a conserved operator, \([K_{\rm R}(\mathbf k),H(\mathbf k)]=0\).
Resolving \(H(\mathbf k)\), and hence \(\rho(\mathbf k)\), into the \(K_{\rm R}=\pm1\) eigensectors gives the spectrum
\begin{align}
    \operatorname{spec}\Gamma(\mathbf k,\theta)
    &=
    \{
        -\delta,
        -\delta,
        +\delta,
        +\delta
    \},
    \\
    \delta(\mathbf k,\theta)
    &=
    \left|
        \cos\!\left[
            \theta-\vartheta_{\rm R}(\mathbf k)
        \right]
    \right|,
    \label{eq:spec_Gamma}
\end{align}
where \(\vartheta_{\rm R}(\mathbf k)\) is the value of \(\theta\) preferred by the state at momentum \(\mathbf k\), and satisfies
\begin{align}
    0
    \leq
    \vartheta_{\rm R}(\mathbf k)
    \leq
    \theta_{\rm s}.
\end{align}
Since \(\Gamma(\theta)\) is block diagonal in momentum space, its spectral gap is set by the minimum of \(\delta(\mathbf k,\theta)\) over the Brillouin zone.
Maximizing this gap gives the optimal angle
\begin{align}
   \theta_{\rm opt}^{\rm R}
    &=
    \frac{\theta_{\rm s}}{2}
    =
    \frac{1}{2}
    \arctan\left(\frac{\lambda_{\rm R}}{A}\right).
    \label{eq:rashba-analytic-angle}
\end{align}
See Appendix~\ref{app:rashba-angle} for a detailed derivation of the optimal angle.

Equation~\eqref{eq:rashba-analytic-angle} determines the optimal angle, $ \theta_{\rm opt}^{\rm R}$ analytically in the translation-invariant system.
To test whether $ \theta_{\rm opt}^{\rm R}$ remains suitable when translation symmetry is broken, we compare it with the angle obtained by maximizing the spectral gap of \(\Gamma(\theta)\) directly in real space.
We denote the eigenvalues of \(\Gamma(\theta)\) by \(\gamma_a(\theta)\), and determine the spectral gap around zero by their smallest absolute value,
\begin{align}
    \Delta_\Gamma(\theta)
    &=
    \min_a
    \left|
        \gamma_a(\theta)
    \right|.
    \label{eq:Gamma-gap}
\end{align}
The corresponding optimal angle is therefore
\begin{align}
    \theta_{\rm opt}
    &=
    \arg\max_\theta
    \Delta_\Gamma(\theta).
    \label{eq:theta_opt}
\end{align}

\subsection{Rashba-Dresselhaus spin-orbit coupling}

Rashba and Dresselhaus spin-orbit coupling arise from two distinct forms of inversion-symmetry breaking.
Rashba coupling is associated with structural inversion asymmetry, for example at an interface, whereas Dresselhaus coupling originates from inversion asymmetry of the underlying material~\cite{zutic2004, winkler2003}.
When both mechanisms are present, the internal hopping matrix in Eq.~\eqref{eq:hopping} becomes
\begin{align}
    t_{jk}
    &=
    t_{jk}^{\rm O}
    +
    t_{jk}^{\rm SO}
    +
    t_{jk}^{\rm R}
    +
    t_{jk}^{\rm D},
    \label{eq:full-hopping}
\end{align}
\begin{align}
    t_{jk}^{\rm D}
    &=
    -\i\lambda_{\rm D}
    \hat{\boldsymbol\gamma}_{jk}\cdot\mathbf s,
        \label{eq:dresselhaus}
\end{align}
where
\begin{align}
    \hat{\boldsymbol\gamma}_{jk}
    &=
    \hat{\mathbf x}\cos\varphi_{jk}
    -
    \hat{\mathbf y}\sin\varphi_{jk}.
    \label{eq:dress-direction}
\end{align}
The Dresselhaus term, Eq.~\eqref{eq:dresselhaus} modifies the spin structure of the hopping by introducing a component along the spin-orbit direction of \(t_{jk}^{\rm SO}\) in Eq.~\eqref{eq:hopping}.
The spin-orbit hopping \(t_{jk}^{\rm SO}\) couples to the spin component along the bond, \(\hat{\boldsymbol\delta}_{jk}\cdot\mathbf s\), whereas Rashba coupling, $ t_{jk}^{\rm R}$, Eq.~\eqref{eq:rashba}, acts along the perpendicular direction \(\hat{\boldsymbol\eta}_{jk}\cdot\mathbf s\), Eq.~\eqref{eq:rashba-direction}.
These two directions satisfy
\begin{align}
    \hat{\boldsymbol\delta}_{jk}
    \cdot
    \hat{\boldsymbol\eta}_{jk}
    &=
    0
\end{align}
for every bond.
The direction of the Dresselhaus term, Eq.~\eqref{eq:dress-direction}, instead satisfies
\begin{align}
    \hat{\boldsymbol\delta}_{jk}
    \cdot
    \hat{\boldsymbol\gamma}_{jk}
    &=
    \cos(2\varphi_{jk}),
\end{align}
and is therefore not generally orthogonal to the original spin-orbit direction.
On an \(x\) bond, the $t_{jk}^{\rm SO}$ and Dresselhaus terms both involve \(s_x\), while the Rashba term involves \(s_y\); on a \(y\) bond, the corresponding components are \(s_y\) and \(s_x\).
The loss of this bondwise orthogonality means that the Dresselhaus term no longer generates the same two-dimensional operator structure that reduced the Rashba problem to the family \(S(\theta)\) in Eq.~\eqref{eq:s_theta}.
The optimal local operator nevertheless remains within the one-parameter family \(S(\theta)\), Eq.~\eqref{eq:s_theta}, because the symmetries of the Hamiltonian strongly restrict the allowed onsite deformations.

For the purpose of the CS-descendant marker, it is sufficient to identify a local Hermitian, time-reversal-odd involution for which \(\Gamma\) remains spectrally gapped.
Any operator satisfying these conditions provides a valid choice of \(S\):The choice of $S$ is therefore not unique.
We take the Rashba family \(S(\theta)\) in Eq.~\eqref{eq:s_theta} as a starting point, and ask how Dresselhaus coupling modifies this ansatz.
To address this question analytically, we consider the translation-invariant square-lattice limit.
In this limit, the Hamiltonian in Eq.~\eqref{eq:model}, with \(W=0\) and the internal hopping matrix in Eq.~\eqref{eq:full-hopping}, is invariant under the twofold rotation \(C_{2z}\), corresponding to a rotation by \(\pi\) about the \(z\) axis.
We therefore restrict the deformations of \(S(\theta)\) to those that respect \(C_{2z}\), while preserving Hermiticity, time-reversal oddness, and the involution condition \(S^2=\id\).
The complete set of infinitesimal deformations satisfying these conditions is derived in Appendix~\ref{app:Dress} and consists of
\begin{equation}
\begin{aligned}
G_1 &= \sin\theta\,\tau_x s_x + \cos\theta\, s_y,\\
G_2 &= \sin\theta\,\tau_x s_y - \cos\theta\, s_x,\\
G_3 &= \tau_x s_z,  ~
G_4 = \tau_z s_x, ~
G_5 = \tau_z s_y.\\
\end{aligned}
\label{eq:allowed-perturbations}
\end{equation}
The remaining tangent direction $G_6=
    \partial_\theta S(\theta)
    =
    \cos\theta\,\tau_y
    -
    \sin\theta\,\tau_zs_z,$ only changes the angle \(\theta\) and therefore does not enlarge the family in Eq.~\eqref{eq:s_theta}.
The only possible deformation which respects the \(C_{2z}\) symmetry is \(G_3=\tau_xs_z\), which results in the two-parameter family
\begin{align}
    S(\theta,\phi)
    &=
    \cos\phi\,S(\theta)
    +
    \sin\phi\,\tau_xs_z.
    \label{eq:two-angle-family}
\end{align}
\begin{figure*}[t]
    \centering
    \includegraphics[width=1\linewidth]{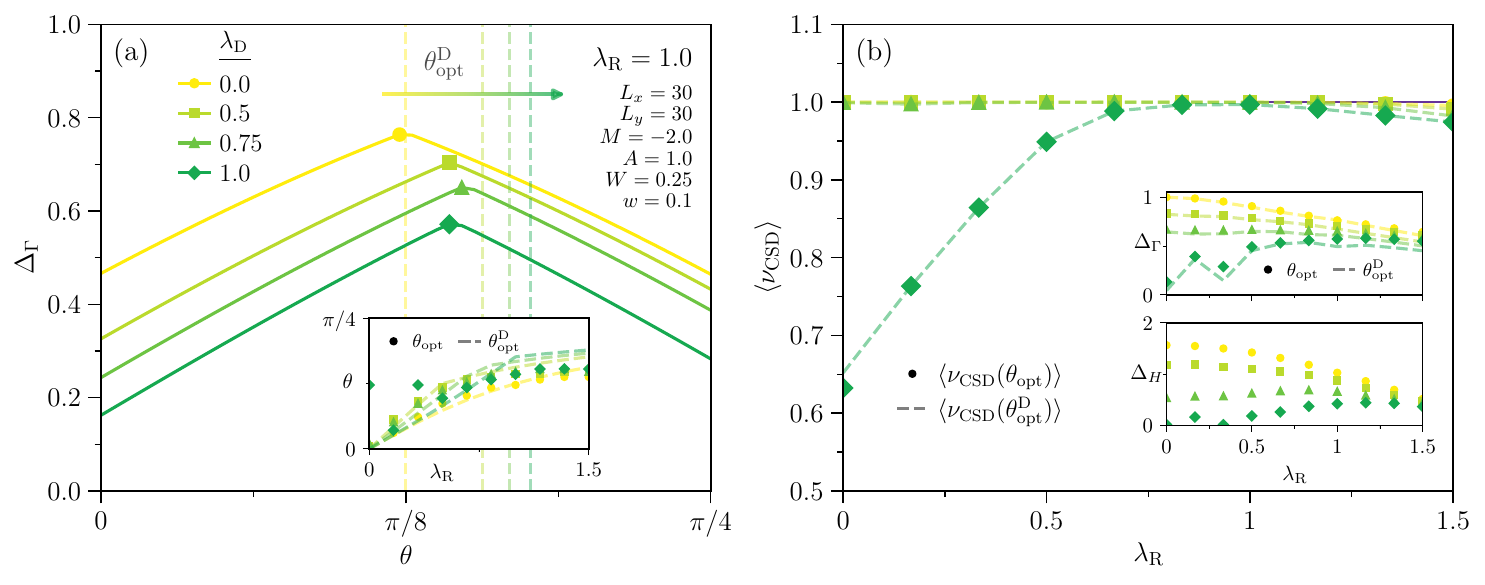}
    \caption{(a) Spectral gap of the operator $\Gamma$, Eq.~(\ref{eq:Gamma}), as a function of the angle $\theta$.
    $\Gamma$ is constructed from the one-parameter family $S(\theta)$, Eq.~(\ref{eq:s_theta}), and the one-particle density matrix of the ground state of the Hamiltonian in Eq.~(\ref{eq:model}), with internal hopping matrix given by Eq.~(\ref{eq:full-hopping}) and defined in an amorphous arrangement with $w=0.1$ constructed from a parent square lattice with linear system sizes $L_x=L_y=30$ and periodic boundary conditions.
    Aside from amorphicity, we use onsite disorder of strength $W=0.25$, and set the parameters $M=-2, A=1$.
    The different solid lines denote different values of the Dresselhaus spin-orbit coupling, while the Rashba contribution is fixed to $\lambda=1$.
    The dashed lines mark the positions of the optimal analytical angle $\theta_{\rm opt}^{\rm D}$ obtained in the limit of translation invariance, Eq.~(\ref{eq:theta-opt-dresselhaus}).
    The inset shows the optimal numerical angle $\theta_{\rm opt}$ (markers) and the optimal analytical angle $\theta_{\rm opt}^{\rm D}$ (dashed lines) as a function of the Rashba coupling.
    The colors refer to the same Dresselhaus coupling strengths as in the main panel.
    (b) Bulk-averaged local Chern-Simons descendant marker for different Dresselhaus couplings $\lambda_{\rm D}$ as a function of the Rashba coupling $\lambda_{\rm R}$ for the ground state of Eq.~(\ref{eq:model}) and same parameters as in (a).
    The value of the marker is obtained through Eq.~(\ref{eq:general-local-marker}) by constructing $\Gamma$ for both the numerical optimal angle $\theta_{\rm opt}$ (markers) and the optimal analytical angle $\theta_{\rm opt}^D$ (dashed lines).
    To calculate the bulk-averaged marker we use periodic boundary conditions in the amorphous arrangement and average over the whole system.
    The data shows a single amorphicity and disorder realization for each value $\lambda_{\rm R}$.
    The upper inset shows the spectral gap of the operator $\Gamma$ as a function of $\lambda_{\rm R}$ constructed from the optimal numerical and analytical angles.
    The lower inset shows the spectral gap of the Hamiltonian as a function of $\lambda_{\rm R}$.
    }
    \label{fig:optimal_angle}
\end{figure*}
The additional direction \(G_3=\tau_xs_z\) does not improve the spectral gap of \(\Gamma\).
Since \(G_3\) anticommutes with \( H_{\rm RD}\), it exchanges occupied and unoccupied states, such that
\begin{align}
    \{\rho,G_3\}
    &=
    G_3.
\end{align}
Using Eq.~\eqref{eq:two-angle-family} in \(\Gamma\), Eq.~\eqref{eq:Gamma}, therefore gives \(\Gamma(\theta,\phi)=\cos\phi\,\Gamma(\theta)\), so the spectral gap
is maximal at \(\phi=0\).
The optimal operator therefore remains within the one-parameter family \(S(\theta)\).

Although structural disorder removes the \(C_{2z}\) symmetry used in the analytic argument, the local coordination and hopping structure of the model remains.
We therefore consider the family \(S(\theta)\) as a candidate for $S$ in the disordered system and test its validity numerically.
Perturbation theory motivates the use of the one-parameter family \(S(\theta)\) even without assuming translation invariance.
For \(\lambda_{\rm D}\ll\lambda_{\rm R}\), the Dresselhaus term acts as a weak perturbation of the model with Rashba spin orbit coupling, so its leading effect is
expected to shift the value of \(\theta\) that maximizes the gap of \(\Gamma\), rather than to require a qualitatively different local operator \(S\).

We invoke translation invariance to analytically determine the optimal angle $\theta$, which maximizes the spectral gap of $\Gamma$.
The Bloch Hamiltonian, $H_{\rm{RD}}(\bold{k})$ retains the form of Eq.~\eqref{eq:block-ham-rashba}, but the Dresselhaus term modifies \(\mathbf h(\mathbf k)\) to
\begin{align}
    \mathbf h(\mathbf k)
    &=
    2
    \begin{pmatrix}
        \lambda_{\rm R}\sin k_y+\lambda_{\rm D}\sin k_x\\
        -\lambda_{\rm R}\sin k_x-\lambda_{\rm D}\sin k_y
    \end{pmatrix}.
\end{align}
With \(\mathbf g(\mathbf k)=2A(\sin k_x,\sin k_y)\), the two vectors now satisfy
\begin{align}
    \mathbf g(\mathbf k)\cdot\mathbf h(\mathbf k)
    &=
    4A\lambda_{\rm D}
    \left(
        \sin^2 k_x-\sin^2 k_y
    \right),
\end{align}
so the orthogonality in Eq.~\eqref{eq:rasha-hg-orthogonal} is lost.
The Bloch Hamiltonian $H_{\rm{RD}}(\bold{k})$ nevertheless admits a conserved sector decomposition, derived in Appendix~\ref{app:Dress}.
The spectrum of \(\Gamma(\mathbf k,\theta)\) retains the form of Eq.~\eqref{eq:spec_Gamma}, but with a modified momentum-dependent angle \(\vartheta(\mathbf k)\).
The eigenvalues are maximal in magnitude when \(\theta=\vartheta(\mathbf k)\), which defines the preferred angle at momentum \(\mathbf k\).
The physical band gap of the Hamiltonian $H_{\rm{RD}}(\bold{k})$ remains open for the parameter range \(0\leq\lambda_{\rm D}<\sqrt{A^2+\lambda_{\rm R}^2}\),in which the preferred angle \(\vartheta(\mathbf k)\) spans the interval \([\vartheta_{\rm min},\vartheta_{\rm max}]\), with
\begin{align}
    \vartheta_{\rm max}
    &=
    \arctan\frac{\lambda_{\rm R}+\lambda_{\rm D}}{A},
    \\
    \vartheta_{\rm min}
    &=
    \arctan
    \frac{\min(0,\lambda_{\rm R}-\lambda_{\rm D})}{A}.
\end{align}
Maximizing the minimum gap of $\Gamma$ over momentum yield the angle
\begin{align}
    \theta_{\rm opt}^{\rm D}
    &=
    \frac{1}{2}
    \left[
        \arctan\frac{\lambda_{\rm R}+\lambda_{\rm D}}{A}
        +
        \arctan
        \frac{\min(0,\lambda_{\rm R}-\lambda_{\rm D})}{A}
    \right].
    \label{eq:theta-opt-dresselhaus}
\end{align}

\subsection{Numerical implementation of the one-parameter family $S(\theta)$}

To demonstrate that the family \(S(\theta)\), Eq.~\eqref{eq:s_theta}, remains a suitable choice without translation symmetry, we consider an amorphous model described by the Hamiltonian in Eq.~\eqref{eq:model} with the internal hopping matrix in Eq.~\eqref{eq:full-hopping}.
We generate the amorphous structure by drawing each site from a Gaussian distribution with standard deviation \(\omega=0.1\), centred on the corresponding site of a square lattice with linear dimensions \(L_x=L_y=30\).
The hopping strength, Eq.~\eqref{eq:hopping}, is \(A=1\), the onsite potential, Eq.~\eqref{eq:onsite}, is \(M=-2\), with onsite disorder strength \(W=0.25\).

Fig~\ref{fig:optimal_angle}(a) shows the gap $\Delta_\Gamma$ of \(\Gamma\) as a function of \(\theta\) for \(\lambda_{\rm D}=0,\,0.5,\,0.75,\) and \(1\), with the Rashba coupling fixed at \(\lambda_{\rm R}=1\).
The gap peaks at an optimal value of \(\theta\), but remains open over a broad range of angles around its maximum, including the analytical optimal angle \(\theta_{\rm opt}^{\rm D}\).
Increasing \(\lambda_{\rm D}\) shifts the maximum of \(\Delta_\Gamma\) towards larger values of \(\theta\), while the gap remains open.
These results illustrate that the local CS-descendant marker does not require a finely tuned choice of \(S(\theta)\):  \(\Gamma\) remains gapped for a broad range of angles, and therefore defines a valid marker through Eq.~\eqref{eq:general-local-marker}.

The inset in Fig.~\ref{fig:optimal_angle}(a) shows the numerically obtained optimal angle \(\theta_{\rm opt}\) and the analytical result \(\theta_{\rm opt}^{\rm D}\) as functions of \(\lambda_{\rm R}\).
The two angles remain close throughout the range of \(\lambda_{\rm R}\) for all four values of \(\lambda_{\rm D}\).
Thus, the optimal angle obtained in the translation-invariant limit provides a robust choice of \(S(\theta)\) even when translation symmetry is broken by structural disorder.

For a single structural and onsite disorder realization, the bulk-averaged CS-descendant marker remains quantized at \(\langle\nu_{\rm CSD}\rangle=1\) throughout the range
\(0\leq\lambda_{\rm R}\lesssim1.5\), \(\lambda_{\rm D}\lesssim2\), for both the analytical and numerically optimized choices of \(S(\theta)\), as shown in Fig.~\ref{fig:optimal_angle}(b).
The deviations from the quantized value occur when the spectrum of $H$ approaches a gap closing (upper inset in Fig.~\ref{fig:optimal_angle}(b)), both in the case of $\lambda_D=2$ and $\lambda_R\simeq1.5$.
These deviations are expected as the spectral gap of $H$ sets the localization length of the one-particle density matrix, yielding an ill-defined CS-descendant marker when $\Delta H\simeq0$.
The lower inset in Fig.~\ref{fig:optimal_angle}(b) shows that the spectral gap of $\Gamma$ remains open across the whole parameter range for both choices $\theta_{\rm opt}$ and $\theta_{\rm opt}^{\rm D}$.

The results shown in Fig.~\ref{fig:optimal_angle} demonstrate that the angle obtained from the translation-invariant limit provides a robust ansatz when translation symmetry is broken.
Although direct optimization can further increase \(\Delta_\Gamma\), it is not required to obtain a well-defined and quantized marker over the parameter range considered here.

\begin{figure}[tb]
    \centering
    \includegraphics[width=\columnwidth]{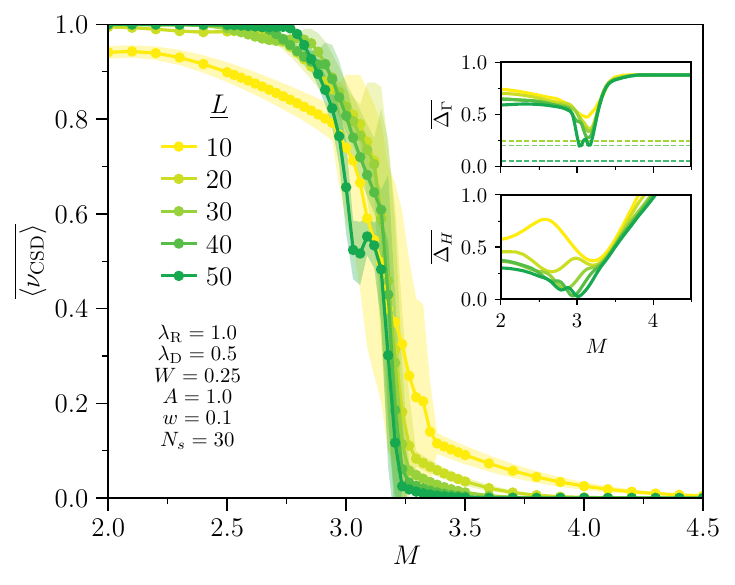}
    \caption{Bulk- and sample-averaged CS-descendant marker as a function of the onsite potential $M$ and for linear system sizes: $L =10,20,30,40,50$.
    The marker is calculated for the ground state of the Hamiltonian in Eq.~(\ref{eq:model}), with internal hopping matrix given by Eq.~(\ref{eq:full-hopping}) and defined in an amorphous arrangement with $w=0.1$ constructed from a parent square lattice with periodic boundary conditions.
    We consider onsite disorder of strength $W=0.25$, and set the parameters $A=1, \lambda_R=1, \lambda_D=0.5$.
    At each data point we average the marker over the whole arrangement of sites and over $N_s=30$ samples of the amorphous and disordered arrangement.
    The shaded contours show the standard deviation of the amorphicity and disorder distribution.
  The CS-descendant marker is calculated from Eq.~(\ref{eq:general-local-marker}) by constructing the operator $\Gamma$, Eq.~(\ref{eq:Gamma}), from the one-parameter family $S(\theta)$, Eq.~(\ref{eq:s_theta}), choosing the angle $\theta \in \{0, \theta_k/2, \theta_k \}$ that maximizes the spectral gap of $\Gamma$.
    The upper inset shows the sample-averaged value of the spectral gap $\Delta_{\Gamma}$ of $\Gamma$ as a function of $M$. 
    The dashed lines mark the absolute minimum $\Delta_{\Gamma}$ obtained in any single amorphicity and disorder realization for each system size.
    The lower inset shows the sample-averaged spectral gap $\Delta_{H}$ of the Hamiltonian as a function of $M$.
    }
    \label{fig:transition}
\end{figure}

Fig.~\ref{fig:transition} shows the CS-descendant marker across a topological transition in an amorphous model with onsite disorder as the onsite potential varies from \(M=2\) to \(M=4.5\).
The other parameters are fixed at \(\lambda_{\rm R}=1\), \(\lambda_{\rm D}=0.5\), \(A=1\), \(W=0.25\), and \(\omega=0.1\).
At each value of \(M\), we average the marker over the amorphous structure and over \(N_s=30\) structural and onsite disorder realizations, with the shaded regions indicating the corresponding standard deviation.

The marker remains quantized on both sides of the transition, with \(\overline{\langle\nu_{\rm CSD}\rangle}=1\) for \(M\lesssim2.75\) and \(\overline{\langle\nu_{\rm CSD}\rangle}=0\) for \(M\gtrsim3.25\).
Increasing the system size sharpens the transition, consistent with the behavior expected in the thermodynamic limit.
Our results are consistent with the clean crystalline limit, where the topological transition occurs at \(M=4\), showing that disorder drives the transition to a smaller value of \(M\).
We evaluate the marker using Eq.~\eqref{eq:general-local-marker} for \(\theta\in\{0,\theta_{\rm opt}^{\rm D}/2, \theta_{\rm opt}^{\rm D}\}\), choosing the angle that gives the largest
spectral gap of \(\Gamma(\theta)\).
This optimization over \(\theta\) ensures that the spectral gap of \(\Gamma\) remains open across the transition, where it becomes more sensitive to the choice of \(S\).

The two insets in Fig.~\ref{fig:transition} show the sample-averaged spectral gaps of $\Gamma$ and the Hamiltonian, as functions of the onsite potential \(M\).
The minimum value of \(\Delta_\Gamma\) across all samples for each system size, remains finite throughout the parameter range, 
As expected, the value of $M$ at which $\Delta_H$ closes coincides with the transition in the main figure.

The numerical results in Figs.~\ref{fig:optimal_angle} and \ref{fig:transition} demonstrate that the family \(S(\theta)\) remains an adequate choice for \(S\) when structural and onsite disorder break translation symmetry.

\subsection{More general choices of \(S\)}

Physical information about the state provides a simple recipe for choosing a well-defined \(S\), as illustrated explicitly for Rashba and Dresselhaus spin-orbit coupling.
To make the prescription for determining \(S\) complete, we nevertheless develop a general numerical procedure that searches over a broader class of local operators, for the exceptional cases where no suitable low dimensional parameter family of $S$ is apparent.

The local CS-descendant marker requires \(\Gamma\) to remain spectrally gapped around zero.
Due to the Hermicity of $\rho$ and $S$, $\rho^2=\rho$, and $S^2=1$,
\begin{equation}
\Gamma^2=1+[\rho,S]^2
    \label{eq:S-auxiliary-square},
\end{equation}
which shows that a choice of \(S\) for which \([S,\rho]\) remains small preserves this gap.
For strongly localized states, any local \(S\) that commutes with the onsite blocks of the one-particle density matrix has a small commutator
\([S,\rho]\):  the onsite blocks dominate \(\rho\), while the remaining intercell blocks are suppressed by localization.
We define \(S\) to act locally,
\begin{align}
    S
    &=
    \bigoplus_{\mathbf r} S_{\mathbf r},
\end{align}
where \(\mathbf r\) labels a unit cell and the local Hilbert-space dimension may vary across the state.
By writing the one-particle density matrix in spatial blocks \(\rho_{\mathbf r\mathbf r'}\), the corresponding blocks of the commutator are
\begin{align}
    [S,\rho]_{\mathbf r\mathbf r'}
    &=
    S_{\mathbf r}\rho_{\mathbf r\mathbf r'}
    -
    \rho_{\mathbf r\mathbf r'}S_{\mathbf r'} ,
    \label{eq:S-rho-local-blocks}
\end{align}
which reduces to \([S_{\mathbf r},\rho_{\mathbf r\mathbf r}]\) for \(\mathbf r=\mathbf r'\).
So choosing each \(S_{\mathbf r}\) to commute with the corresponding onsite block, therefore removes the onsite contribution in Eq.~\eqref{eq:S-rho-local-blocks} exactly, and the remaining contributions come entirely from blocks \(\rho_{\mathbf r\mathbf r'}\) that connect different unit cells.
By defining \(\rho=\rho_{\rm on}+\rho_{\rm off}\), with \(\rho_{\rm on}=\bigoplus_{\mathbf r}\rho_{\mathbf r\mathbf r}\), the commutator \([S,\rho]=[S,\rho_{\rm off}]\), such that \(\|[S,\rho]\|\leq2\|\rho_{\rm off}\|\).
Strong localization suppresses \(\|\rho_{\rm off}\|\), in turn suppressing \([S,\rho]\), which guarantees through Eq.~\eqref{eq:S-auxiliary-square} that \(\Gamma\) has a large spectral gap around zero. 

To construct the local operators \(S_{\mathbf r}\) that commute with the onsite blocks \(\rho_{\mathbf r\mathbf r}\), we consider a four-dimensional onsite Hilbert space formed by two orbital and two spin degrees of freedom,  as in the other examples in this section.
Time-reversal symmetry enforces the eigenvectors of \(\rho_{\mathbf r\mathbf r}\) to come in Kramers pairs.
Within each pair, \(\rho_{\mathbf r\mathbf r}\) is proportional to the identity, so any choice of axis commutes with the onsite block.
Assigning eigenvalues \(+1\) and \(-1\) along this axis gives a Hermitian involution that is odd under time reversal.
The most general local choice of this form is
\begin{align}
    S_{\mathbf r}
    &=
    \sum_{K=1}^{2}
    W_{\mathbf rK}
    (\hat{\mathbf n}_{\mathbf rK}\cdot\boldsymbol\sigma)
    W_{\mathbf rK}^{\dagger},
    &
    |\hat{\mathbf n}_{\mathbf rK}|
    &=
    1,
    \label{eq:S-local-kramers-ansatz}
\end{align}
where the columns of \(W_{\mathbf rK}\) form an orthonormal Kramers pair of eigenvectors of \(\rho_{\mathbf r\mathbf r}\), and \(\boldsymbol\sigma\) acts within that pair.
Every choice of the unit vectors \(\hat{\mathbf n}_{\mathbf rK}\) satisfies \(S_{\mathbf r}^{\dagger}=S_{\mathbf r}\), \(S_{\mathbf r}^{2}=1\), \(\mathcal T S_{\mathbf r}\mathcal T^{-1}=-S_{\mathbf r}\), and \([S_{\mathbf r},\rho_{\mathbf r\mathbf r}]=0\).
Hence, in the strongly localized limit, the directions \(\hat{\mathbf n}_{\mathbf rK}\) are largely unrestricted: every member of this local Kramers-pair family eliminates the onsite contribution to \([S,\rho]\), while localization suppresses the remaining intercell contribution.
Any choice within the family in Eq.~\eqref{eq:S-local-kramers-ansatz} therefore keeps \([S,\rho]\) small and produces a large auxiliary gap.
Strong structural or onsite disorder therefore provides a particularly simple setting for determining \(S\), even simpler than for the analytically solvable translation-invariant case.

The local parametrization in Eq.~\eqref{eq:S-local-kramers-ansatz} does not rely on strong localization; only the freedom to choose the directions \(\hat{\mathbf n}_{\mathbf rK}\) arbitrarily does.
As the localization length increases, the intercell blocks contribute more strongly to \([S,\rho]\), so the orientations \(\hat{\mathbf n}_{\mathbf rK}\) become important for maintaining the spectral gap of \(\Gamma\).
For an exponentially localized one-particle density matrix,
\begin{align}
    \|\rho_{\mathbf r\mathbf r'}\|
    &\leq
    C e^{-|\mathbf r-\mathbf r'|/\xi},
\end{align}
the same local Kramers-pair family therefore provides a natural parametrization of \(S\) beyond the strongly localized limit.
The vectors \(\hat{\mathbf n}_{\mathbf rK}\) can be used as variational degrees of freedom and in Appendix~\ref{app:S-optimization}, we discuss how determine them numerically by maximizing the spectral gap of \(\Gamma\).
We show that the initial optimization maps onto a classical spin-glass problem on the original lattice, with two three-component unit vectors per site, and Newton's method subsequently refines the resulting configuration.

\section{Discussion}

We have derived the local Chern-Simons descendant marker for time-reversal-invariant states in two dimensions directly from the one-particle density matrix of a given state.
The marker is obtained by dimensional reduction of the three-dimensional local Chern-Simons marker and introduces a local Hermitian involution \(S\) that is odd under time reversal.
When \(S\) commutes with the one-particle density matrix, the occupied subspace separates into the \(S=\pm1\) eigensectors and the CS-descendant marker reduces to half the difference of their local Chern markers, modulo two.
A conserved physical spin therefore reproduces the spin-Chern marker as a special case: more general choices of \(S\) extend the the use of the marker to states without spin conservation.

The local CS-descendant marker remains well defined even when \([S,\rho]\neq0\), provided that the operator \(\Gamma=\{\rho,S\}-S\) has a spectral gap around zero.
The spectral gap of \(\Gamma\) therefore guarantees that the marker is well defined and provides a practical criterion for choosing \(S\).
The freedom in \(S\) is important because it allows the marker to describe states in which physical spin is not conserved.
We have demonstrated that this freedom does not require an unconstrained search over local operators.
On the contrary, the physical structure of the state often identifies a small family of natural candidates for \(S\).

We identify choices of \(S\) that keep \(\Gamma\) spectrally gapped across models with progressively stronger spin mixing, considering both structural and potential disorder, and  stress that a spin-orbital symmetry gives an exactly conserved \(S\) even when physical spin is not conserved.
This becomes particularly relevant in amorphous matter, where the local spin frame may vary with the surrounding structure: the local spin rotations turn the same physical symmetry into a spatially varying operator \(S(\mathbf r)\), without requiring a new choice of \(S\).

We introduce both Rashba and Dresselhaus spin-orbit coupling to demonstrate how the spin-orbital structure reduces the search for \(S\) to a simple one-parameter family \(S(\theta)\), despite the loss of exact conservation.
We obtain the angle that maximizes the spectral gap of \(\Gamma\) analytically in the translation-invariant limit.
The analytic choice of \(S\) remains effective beyond the translation-invariant setting in which it was derived.
For both structural and scalar potential disorder, we find that the crystalline value of \(\theta\) continues to keep \(\Gamma\) strongly gapped, while a broad interval of nearby angles remains equally admissible, making suitable choices of \(S\) readily accessible.
The same low-dimensional choice of \(S\) therefore survives the loss of translation symmetry without fine tuning.

Our results suggest a practical route for identifying \(S\) in general: determine a suitable choice in a translation-invariant reference problem, check the corresponding gap in $\Gamma$ for the given state.
If the spectral gap of \(\Gamma\) remains large, the reference value of \(S\) provides the choice needed to evaluate the local \(\mathbb{Z}_2\) marker and characterize the topology of the state.
To cover states whose physical structure does not suggest such a low-dimensional family, we also provide a general numerical procedure for determining \(S\) using the locality of the one-particle density matrix.
The systematic prescription for determining \(S\) makes the local Chern-Simons descendant marker a practical tool for characterizing topology in real space.

Although we have derived the marker explicitly for class AII, the same local \(\mathbb Z_2\) marker also characterizes two-dimensional phases in class DIII.
For a class-DIII state, particle-hole symmetry requires the relevant one-particle density matrix to be the full Bogoliubov-de Gennes one-particle density matrix \(\varrho\), including the anomalous pairing correlations.
The local \(\mathbb Z_2\) marker therefore characterizes both time-reversal-invariant insulators and time-reversal-invariant superconductors.
The dimensional reduction used to obtain the two-dimensional marker applies in every even spatial dimension and yields local markers for all \(\mathbb Z_2\)-classified Altland-Zirnbauer phases in even dimensions.

The local Chern-Simons descendant marker belongs to a broader class of local topological markers formulated directly from the state, which also includes the Chern, chiral, and Chern-Simons markers.
These state-based topological markers provide a direct local interpretation of topology in Gaussian and near-Gaussian states, with or without a specified parent Hamiltonian.
This is particularly valuable when the state itself is the natural object to characterize, for example for localized mid-spectrum states or in settings without a conventional Hamiltonian spectral gap.
The spatial profiles of the state based local markers, and their dependence on coarse-graining, reveal both how topology is organized within a state and the length scale on which a local topological phase becomes well defined.
This spatial information can be used to distinguish whether finite-size deviations from a well-defined phase arise from finite regions of another phase or from a more homogeneous state with a large topological length scale.

The present work demonstrates that the local characterization of states is practically feasible for two-dimensional time-reversal-invariant $\mathbb Z_2$ states, extending the framework of the local state-based markers to one of the most physically relevant classes of topological matter.
Looking ahead, the spatial information provided by the local Chern-Simons descendant marker could be used to study the mechanisms of topological transitions, particularly in disordered and amorphous systems.
Their spatial information, for example, offers a way to investigate whether a transition proceeds through the formation, growth, and connectivity of domains with different topology, or through a more homogeneous divergence of the topological length scale.
The CS-descendant marker,  as well as all the other local markers in the family of state-based markers, thus provide tools for exploring the physics of topological phases and their transitions, beyond determining which phase a state belongs to.

\section{Acknowledgments}
This work was supported by a Simons Investigator Award (Grant No. 511029).
M.F.M and T.K.K are funded by the Wenner-Gren foundations.
The work of J.D.H. was partly performed at Donostia International Physics Center (DIPC), and J.D.H acknowledges financial support from the DIPC through the HER (Hosting Excellent Researchers) initiative. 
The authors used AI tools from OpenAI and Anthropic for support with code development and language editing. 
The authors take full responsibility for the content of this publication.

\appendix
\section{Dimensional reduction of the local Chern-Simons marker}
\label{app:dimensional-reduction}

We derive the dimensional reduction of the three-dimensional class-CII marker to a local \(\mathbb Z_2\) marker for a two-dimensional state in class AII.

The extension to the auxiliary third dimension is
\begin{align}
    \widetilde\rho(k_z)
    &=
    \frac{1}{2}
    \Big[
        \id
        -
        \sin k_z\,(\id-2\rho)\otimes\tau_x
        -
        \cos k_z\,\id\otimes\tau_z
    \Big],
    \label{eq:supp-bott}
\end{align}
where the Pauli matrices \(\tau_a\), \(a=x,y,z\), act in the two-dimensional auxiliary space, and $k_z$ is the auxiliary dimension.
Since \(\rho^2=\rho\),  \((\id-2\rho)^2=\id\), and \(\{\tau_x,\tau_z\}=0\), we have
\begin{align}
    \widetilde\rho(k_z)^2
    &=
    \frac{1}{4}
    \Big[
        \id
        -
        2\sin k_z\,(\id-2\rho)\otimes\tau_x
        -
        2\cos k_z\,\id\otimes\tau_z
        \notag\\
        &\qquad
        +
        \sin^2 k_z\,\id
        +
        \cos^2 k_z\,\id
    \Big]
    \notag\\
    &=
    \widetilde\rho(k_z).
\end{align}
Let \(\lambda=\pm\) label the auxiliary two-dimensional space and define the combined local index
\begin{align}
    \beta
    &=
    (\alpha,\lambda),
\end{align}
where \(\alpha\) labels the internal degrees of freedom of the original two-dimensional state.
The three-dimensional local Chern-Simons marker in class CII is
\begin{align}
    \nu_{\rm CS}(\mathbf r)
    &=
    -\frac{8\pi\i}{3}
    \int_0^{2\pi}\frac{dk_z}{2\pi}
    \sum_\beta\varepsilon^{abc}
    \notag\\
    &\quad\times
    \left[
        \widetilde\rho\widetilde S X_a
        \widetilde\rho X_b
        \widetilde\rho X_c
        \widetilde\rho
    \right]_{(\mathbf r,\beta),(\mathbf r,\beta)}
    \mod 2 ,
    \label{eq:supp-CS-marker}
\end{align}
where \(a,b,c\in\{1,2,3\}\),
\(X_1\) and \(X_2\) are the physical position operators, and $X_3=
    \i\partial_{k_z}.$
The operator
\begin{align}
    \widetilde S
    &=
    S\otimes\tau_y ,
    \label{eq:supp-Stilde}
\end{align}
where $S$ obeys the conditions:
\begin{align}
    S^\dagger &= S,
    &
    S^2 &= 1,
    &
    TS^*T^\dagger &= -S .
    \label{eq:supp-S-local-conditions}
\end{align}
and 
\begin{align}
    [S,\rho]
    &=
    0 .
    \label{eq:supp-S-commuting}
\end{align}
The time-reversal condition together with \(S^2=1\) pairs the
\(S=+1\) and \(S=-1\) eigenspaces and therefore also implies
\(\Tr S=0\).
The extended projector obeys the chiral condition
\begin{align}
    \widetilde S\widetilde\rho(k_z)\widetilde S
    &=
    \id-\widetilde\rho(k_z),
    \label{eq:supp-rho-chiral}
\end{align}
or, equivalently,
\begin{align}
    \{\widetilde S,\widetilde\rho(k_z)\}
    &=
    \widetilde S .
\end{align}

Since only one of the three directions is auxiliary, the
antisymmetric sum can be separated according to the position of
\(X_3=\i\partial_{k_z}\).
Writing \(i,j\in\{1,2\}\), this gives
\begin{align}
    \varepsilon^{abc}
    \widetilde\rho\widetilde S X_a
    \widetilde\rho X_b
    \widetilde\rho X_c
    \widetilde\rho&=
    \i\varepsilon^{ij}
    \Big(
        \widetilde\rho\widetilde S X_i
        \widetilde\rho X_j
        \widetilde\rho\partial_{k_z}\widetilde\rho
    \notag\\
    &
        -
        \widetilde\rho\widetilde S X_i
        \widetilde\rho\partial_{k_z}\widetilde\rho
        X_j\widetilde\rho
        +
        \widetilde\rho\widetilde S
        \partial_{k_z}\widetilde\rho
        X_i\widetilde\rho X_j\widetilde\rho
    \Big).
    \label{eq:supp-three-terms}
\end{align}

For brevity, define
\begin{align}
    s&=\sin k_z,
    &
    c&=\cos k_z .
\end{align}
We separate the extended projector into its identity and non-identity
parts,
\begin{align}
    \widetilde\rho
    &=
    \frac{1}{2}\id+\delta\widetilde\rho,
    \\
    \delta\widetilde\rho
    &=
    -\frac{1}{2}
    \Big[
        s\,(\id-2\rho)\otimes\tau_x
        +
        c\,\id\otimes\tau_z
    \Big].
    \label{eq:supp-rho-nonidentity}
\end{align}
Its derivative is
\begin{align}
    \partial_{k_z}\widetilde\rho
    =
    \partial_{k_z}\delta\widetilde\rho
    &=
    -\frac{1}{2}
    \Big[
        c\,(\id-2\rho)\otimes\tau_x
        -
        s\,\id\otimes\tau_z
    \Big].
    \label{eq:supp-rho-derivative}
\end{align}
Within the antisymmetrized product in the marker, all contributions in
which an undifferentiated factor of \(\widetilde\rho\) supplies its
identity part vanish.
Each undifferentiated projector in
Eq.~\eqref{eq:supp-three-terms} can therefore be replaced by
\(\delta\widetilde\rho\).
Together with Eq.~\eqref{eq:supp-rho-derivative}, each of the three
undifferentiated projector factors and the derivative
\(\partial_{k_z}\widetilde\rho\) therefore contributes either
\(\tau_x\) or \(\tau_z\) in the auxiliary space, while
\(\widetilde S\) contributes one \(\tau_y\).
Resolving the composite index \(\beta=(\alpha,\lambda)\), the auxiliary
sum contains factors of the form
\begin{align}
    \sum_{\lambda=\pm}
    \left[
        \tau_y
        \tau_{a_1}\tau_{a_2}
        \tau_{a_3}\tau_{a_4}
    \right]_{\lambda\lambda},
    \qquad
    a_n\in\{x,z\}.
    \label{eq:supp-tau-trace}
\end{align}
The Pauli-matrix algebra makes this sum nonzero only when
\begin{align}
    \tau_{a_1}\tau_{a_2}
    \tau_{a_3}\tau_{a_4}
    &\propto
    \tau_y .
\end{align}
Hence the surviving terms contain an odd number of
\(\tau_x\) factors and an odd number of \(\tau_z\) factors.

For the surviving terms in Eq.~\eqref{eq:supp-three-terms}, the
derivative can be combined with an adjacent factor of
\(\delta\widetilde\rho\).
Using \(\rho^2=\rho\), so that
\((\id-2\rho)^2=\id\), together with
\(\tau_x^2=\tau_z^2=\id\), we obtain
\begin{align}
    \delta\widetilde\rho\,
    \partial_{k_z}\widetilde\rho
    &=
    \frac{1}{4}
    \Big[
        sc\,\id\otimes\id
        -
        s^2(\id-2\rho)\otimes\tau_x\tau_z
        \notag\\
    &\qquad\quad
        +
        c^2(\id-2\rho)\otimes\tau_z\tau_x
        -
        sc\,\id\otimes\id
    \Big]
    \notag\\
    &=
    \frac{1}{4}
    \Big[
        -s^2(\id-2\rho)\otimes\tau_x\tau_z
        +
        c^2(\id-2\rho)\otimes\tau_z\tau_x
    \Big]
    \notag\\
    &=
    \frac{\i}{4}
    (\id-2\rho)\otimes\tau_y .
    \label{eq:supp-rhodrho}
\end{align}
Evaluating the auxiliary sum for all three terms in
Eq.~\eqref{eq:supp-three-terms} then gives
\begin{align}
    &\sum_{\lambda=\pm}
    \varepsilon^{abc}
    \Big[
        \widetilde\rho\widetilde S X_a
        \widetilde\rho X_b
        \widetilde\rho X_c
        \widetilde\rho
    \Big]_{\mathbf r\alpha\lambda}=
    \notag\\
    &=
    \frac{3}{8}s^2\varepsilon^{ij}
    \left[
        (\id-2\rho) S X_i
        (\id-2\rho) X_j
        (\id-2\rho)
    \right]_{\mathbf r\alpha}
    \notag\\
    &+
    \frac{1}{8}c^2\varepsilon^{ij}
    \Big[
        S\big(
            X_iX_j(\id-2\rho)
            +
            X_i(\id-2\rho)X_j
            +    \notag\\
    &\hspace{4cm}
            (\id-2\rho)X_iX_j
        \big)
    \Big]_{\mathbf r\alpha}.
    \label{eq:supp-full-tau-trace}
\end{align}
The terms proportional to \(c^2\) vanish under antisymmetrization:
the terms containing \(X_iX_j\) vanish because
\([X_i,X_j]=0\), while the remaining term vanishes using
\([S,X_i]=0\) and cyclicity of the local trace.
Using \([S,\rho]=0\), \(S^2=1\), and \([S,X_i]=0\), we can also
perform the remaining antisymmetrization in the physical directions.
Expanding the three factors of \((\id-2\rho)\), all terms containing
fewer than three factors of \(\rho\) either vanish individually or
cancel pairwise.
The auxiliary trace therefore reduces to
\begin{align}
    &\sum_{\lambda=\pm}
    \varepsilon^{abc}
    \Big[
        \widetilde\rho\widetilde S X_a
        \widetilde\rho X_b
        \widetilde\rho X_c
        \widetilde\rho
    \Big]_{\mathbf r\alpha\lambda}
    \notag\\
    &\qquad=
    -3\sin^2 k_z\,
    \varepsilon^{ij}
    \left[
        \rho S X_i
        \rho S X_j
        \rho S
    \right]_{\mathbf r\alpha}.
    \label{eq:supp-tau-trace-reduced}
\end{align}
Substituting Eq.~\eqref{eq:supp-tau-trace-reduced} into
Eq.~\eqref{eq:supp-CS-marker} gives
\begin{align}
    \nu_{\rm CS}(\mathbf r)
    &=
    8\pi\i
    \int_{\mathcal D_{\rm B}}
    \frac{dk_z}{2\pi}\,
    \sin^2 k_z
    \notag\\
    &\quad\times
    \sum_\alpha
    \varepsilon^{ij}
    \left[
        \rho S X_i
        \rho S X_j
        \rho S
    \right]_{\mathbf r\alpha}
    \mod 2 .
    \label{eq:supp-intermediate-rho-marker}
\end{align}
The remaining \(k_z\) dependence is contained entirely in the scalar
integral
\begin{align}
    \mathcal I_{\rm B}
    &\equiv
    \int_{\mathcal D_{\rm B}}
    \frac{dk_z}{2\pi}\,
    \sin^2 k_z ,
    \label{eq:supp-bott-integral}
\end{align}
where \(\mathcal D_{\rm B}\) denotes the fundamental domain of the
auxiliary Bott extension.
The remaining local Chern-Simons descendant marker in two dimension is
\begin{align}
    \nu_{\rm CSD}(\mathbf r)
    &=
    8\pi\i\,\mathcal I_{\rm B}
    \sum_\alpha
    \varepsilon^{ij}
    \left[
        \rho S X_i
        \rho S X_j
        \rho S
    \right]_{\mathbf r\alpha}
    \mod 2 .
    \label{eq:supp-marker-bott-normalization}
\end{align}

The value of \(\mathcal I_{\rm B}\) depends on the fundamental domain used to parametrize the Bott extension.
%
The commuting condition in Eq.~\eqref{eq:supp-S-commuting} provides a direct interpretation of the reduced marker and fixes its normalization.
Define the projectors onto the two eigenspaces of \(S\) by
\begin{align}
    P_\pm
    &=
    \frac{1\pm S}{2},
    &
    \rho_\pm
    &=
    P_\pm\rho P_\pm .
\end{align}
Since \([\rho,S]=0\),
\begin{align}
    \rho
    &=
    \rho_+
    +
    \rho_-,
    &
    \rho S
    &=
    \rho_+
    -
    \rho_- .
\end{align}
The projectors \(P_\pm\) are local and therefore commute with the
position operators.
Mixed-sector products consequently vanish, giving
\begin{align}
    &\rho S X_i
    \rho S X_j
    \rho S
    \notag\\
    &\qquad=
    \rho_+X_i\rho_+X_j\rho_+
    -
    \rho_-X_i\rho_-X_j\rho_- .
    \label{eq:supp-sector-difference}
\end{align}
The local Chern markers of the two sectors are
\begin{align}
    \mathcal C_\pm(\mathbf r)
    &=
    2\pi\i
    \sum_\alpha
    \varepsilon^{ij}
    \left[
        \rho_\pm X_i
        \rho_\pm X_j
        \rho_\pm
    \right]_{\mathbf r\alpha}.
    \label{eq:supp-sector-chern-markers}
\end{align}
Substituting Eq.~\eqref{eq:supp-sector-difference} into
Eq.~\eqref{eq:supp-marker-bott-normalization} gives
\begin{align}
    \nu_{\rm CS}(\mathbf r)
    &=
    4\mathcal I_{\rm B}
    \left[
        \mathcal C_+(\mathbf r)
        -
        \mathcal C_-(\mathbf r)
    \right]
    \mod 2 .
    \label{eq:supp-marker-sector-general}
\end{align}
Time reversal exchanges the two sectors.
In a translation-invariant bulk their Chern numbers therefore satisfy
\begin{align}
    C_-
    &=
    -C_+ ,
\end{align}
so the bulk marker becomes
\begin{align}
    \nu_{\rm CS}
    &=
    8\mathcal I_{\rm B}\,C_+
    \mod 2 .
    \label{eq:supp-marker-sector-bulk}
\end{align}
The minimal nontrivial time-reversal-invariant phase has
\(\lvert C_+\rvert=1\).
Normalizing the corresponding Bott representative to the minimal
nontrivial integer value fixes
\begin{align}
    8\mathcal I_{\rm B}
    &=
    1,
    &
    \mathcal I_{\rm B}
    &=
    \frac{1}{8}.
    \label{eq:supp-bott-normalization}
\end{align}
The fundamental domain and normalization of the auxiliary
\(k_z\) integration must therefore be chosen consistently with
\(\mathcal I_{\rm B}=1/8\).
Equation~\eqref{eq:supp-marker-bott-normalization} then reduces to
\begin{align}
    \nu_{\rm CSD}(\mathbf r)
    &=
    \pi\i
    \sum_\alpha
    \varepsilon^{ij}
    \left[
        \rho S X_i
        \rho S X_j
        \rho S
    \right]_{\mathbf r\alpha}
    \mod 2 ,
    \label{eq:supp-conserved-S-marker}
\end{align}
or, equivalently,
\begin{align}
    \nu_{\rm CSD}(\mathbf r)
    &=
    \frac{1}{2}
    \left[
        \mathcal C_+(\mathbf r)
        -
        \mathcal C_-(\mathbf r)
    \right]
    \mod 2 .
    \label{eq:supp-marker-sector-form}
\end{align}
In a translation-invariant bulk, time reversal exchanges the two
sectors and their Chern numbers satisfy \(C_-=-C_+\).
The bulk invariant therefore reduces to
\begin{align}
    \nu_{\rm CSD}
    &=
    \frac{1}{2}
    \left(
        C_+-C_-
    \right)
    \mod 2
    \notag\\
    &=
    C_+
    \mod 2 .
\end{align}
This result does not require \(S\) to represent physical spin.
A conserved physical spin operator is a special case, for which the two sectors coincide with the spin-up and spin-down occupied sectors and Eq.~\eqref{eq:supp-marker-sector-form} reduces to the familiar spin-Chern expression for the quantum spin Hall invariant.
The local two dimensional CS-descendant marker therefore follows from dimensional reduction of the local Chern-Simons marker without identifying \(S\) with physical spin.

\section{Analytic determination of \(S\) with Rashba spin-orbit coupling}
\label{app:rashba-angle}

The translation-invariant square-lattice Hamiltonian with Rashba
spin-orbit coupling is
\begin{align}
    H(\mathbf k)
    &=
    d(\mathbf k)\tau_z
    +
    2A
    \left(
        \sin k_x\,\tau_xs_x
        +
        \sin k_y\,\tau_xs_y
    \right)
    \nonumber\\
    &\quad
    +
    2\lambda_{\rm R}
    \left(
        \sin k_y\,s_x
        -
        \sin k_x\,s_y
    \right),
    \label{eq:app-rashba-H-full}
\end{align}
with
\begin{align}
    d(\mathbf k)
    &=
    M+2A(\cos k_x+\cos k_y).
\end{align}
Here \(\tau_i\) act on the orbital degrees of freedom and \(s_i\) on
physical spin.
The term proportional to \(A\) is the original BHZ spin-orbit
coupling, while \(\lambda_{\rm R}\) controls the Rashba coupling.
In the absence of Rashba coupling, the local operator
\(S_0=\tau_zs_z\) commutes with the Hamiltonian.
It is useful to separate the two spin-dependent contributions as
\begin{align}
    H_{\rm SO}(\mathbf k)
    &=
    2A
    \left(
        \sin k_x\,\tau_xs_x
        +
        \sin k_y\,\tau_xs_y
    \right),
    \\
    H_{\rm R}(\mathbf k)
    &=
    2\lambda_{\rm R}
    \left(
        \sin k_y\,s_x
        -
        \sin k_x\,s_y
    \right).
\end{align}
For \(\lambda_{\rm R}=0\), the operator
\begin{align}
    S_0
    &=
    \tau_zs_z
\end{align}
commutes with the Hamiltonian.
The Rashba term breaks this conservation.
Using
\([s_z,s_x]=2\i s_y\) and
\([s_z,s_y]=-2\i s_x\), its commutator with \(S_0\) is
\begin{align}
    [S_0,H_{\rm R}]
    &=
    4\i\lambda_{\rm R}
    \left(
        \sin k_x\,\tau_zs_x
        +
        \sin k_y\,\tau_zs_y
    \right).
    \label{eq:app-rashba-comm-S0}
\end{align}
The same internal matrix structure is generated by commuting
\(\tau_y\) with the original BHZ spin-orbit term,
\begin{align}
    [\tau_y,H_{\rm SO}]
    &=
    -4\i A
    \left(
        \sin k_x\,\tau_zs_x
        +
        \sin k_y\,\tau_zs_y
    \right).
    \label{eq:app-rashba-comm-tauy}
\end{align}
Consequently,
\begin{align}
    [S_0,H_{\rm R}]
    &=
    -\frac{\lambda_{\rm R}}{A}
    [\tau_y,H_{\rm SO}].
    \label{eq:app-rashba-comm-relation}
\end{align}
The Rashba perturbation therefore selects \(\tau_y\) as the local
direction in which \(S_0\) should be deformed.

The two operators \(S_0=\tau_zs_z\) and \(\tau_y\) are both Hermitian,
traceless, and odd under time reversal.
They also satisfy
\begin{align}
    S_0^2
    &=
    \tau_y^2
    =
    \id,
    &
    \{S_0,\tau_y\}
    &=
    0.
\end{align}
A normalized linear combination of these two operators therefore
remains an admissible local involution.
Writing the two coefficients as \(\cos\theta\) and \(\sin\theta\)
gives
\begin{align}
    S(\theta)
    &=
    \cos\theta\,\tau_zs_z
    +
    \sin\theta\,\tau_y.
    \label{eq:app-rashba-S}
\end{align}
The problem is thus reduced to determining the angle \(\theta\) that maximizes the spectral gap of
\begin{align}
    \Gamma(\theta)
    &=
    \{\rho,S(\theta)\}
    -
    S(\theta).
    \label{eq:app-rashba-Gamma}
\end{align}

Translation invariance allows this problem to be resolved independently
at each crystal momentum.
To expose the spin structure of Eq.~\eqref{eq:app-rashba-H-full}, define
\begin{align}
    r(\mathbf k)
    &=
    \sqrt{\sin^2 k_x+\sin^2 k_y},
    \\
    s_{\parallel}(\mathbf k)
    &=
    \frac{
        \sin k_x\,s_x+\sin k_y\,s_y
    }{
        r(\mathbf k)
    },
    \\
    s_{\perp}(\mathbf k)
    &=
    \frac{
        \sin k_y\,s_x-\sin k_x\,s_y
    }{
        r(\mathbf k)
    }.
\end{align}
For \(r(\mathbf k)\neq0\), these operators satisfy
\(s_{\parallel}^2=s_{\perp}^2=\id\) and
\(\{s_{\parallel},s_{\perp}\}=0\).
The Hamiltonian then takes the compact form
\begin{align}
    H(\mathbf k)
    &=
    d(\mathbf k)\tau_z
    +
    a(\mathbf k)\tau_xs_{\parallel}(\mathbf k)
    +
    b(\mathbf k)s_{\perp}(\mathbf k),
    \label{eq:app-rashba-H}
    \\
    a(\mathbf k)
    &=
    2A r(\mathbf k),
    \qquad
    b(\mathbf k)
    =
    2\lambda_{\rm R}r(\mathbf k).
\end{align}
The two spin-dependent terms therefore point along orthogonal
directions in spin space, while their relative magnitude,
\begin{align}
    \frac{b(\mathbf k)}{a(\mathbf k)}
    &=
    \frac{\lambda_{\rm R}}{A},
    \label{eq:app-rashba-ratio}
\end{align}
is independent of momentum.

At half filling, and provided the physical band gap remains open,
we write
\begin{align}
    Q(\mathbf k)
    &=
    1-2\rho(\mathbf k)
    =
    \operatorname{sgn}H(\mathbf k),
    \\
    \Gamma(\mathbf k,\theta)
    &=
    -\frac{1}{2}
    \left\{
        Q(\mathbf k),S(\theta)
    \right\}.
    \label{eq:app-rashba-Gamma}
\end{align}
The operator
\begin{align}
    K(\mathbf k)
    &=
    \tau_zs_{\perp}(\mathbf k)
\end{align}
commutes with \(H(\mathbf k)\) and satisfies \(K^2=1\).
Indeed, squaring Eq.~\eqref{eq:app-rashba-H} gives
\begin{align}
    H^2(\mathbf k)
    &=
    a^2+d^2+b^2
    +
    2db\,K.
\end{align}
The \(K=\pm1\) sectors therefore have positive energy scales
\begin{align}
    E_{\pm}
    &=
    \sqrt{a^2+(d\pm b)^2}.
    \label{eq:app-rashba-Epm}
\end{align}
With the projectors
\(\Pi_\pm=(1\pm K)/2\), the flattened Hamiltonian takes the form
\begin{align}
    Q
    &=
    H
    \left(
        \frac{\Pi_+}{E_+}
        +
        \frac{\Pi_-}{E_-}
    \right).
    \label{eq:app-rashba-Q}
\end{align}

We now evaluate Eq.~\eqref{eq:app-rashba-Gamma} for
\begin{align}
    S(\theta)
    &=
    \cos\theta\,\tau_zs_z
    +
    \sin\theta\,\tau_y.
\end{align}
Using Eq.~\eqref{eq:app-rashba-Q} and the Pauli algebra gives
\begin{align}
    \Gamma^2(\mathbf k,\theta)
    &=
    \delta^2(\mathbf k,\theta)\,\id,
    \\
    \delta^2(\mathbf k,\theta)
    &=
    \frac{1}{2}
    \left[
        1
        +
        \frac{a^2+d^2-b^2}{E_+E_-}\cos 2\theta
    \right.
    \nonumber\\
    &\hspace{2.0cm}\left.
        +
        \frac{2ab}{E_+E_-}\sin 2\theta
    \right].
    \label{eq:app-rashba-delta}
\end{align}
The identity
\begin{align}
    (E_+E_-)^2
    &=
    \left(a^2+d^2-b^2\right)^2
    +
    4a^2b^2
\end{align}
allows us to define a momentum-dependent angle
\(\vartheta(\mathbf k)\) through
\begin{align}
    \cos 2\vartheta(\mathbf k)
    &=
    \frac{a^2+d^2-b^2}{E_+E_-},
    \\
    \sin 2\vartheta(\mathbf k)
    &=
    \frac{2ab}{E_+E_-}.
\end{align}
Equation~\eqref{eq:app-rashba-delta} then reduces to
\begin{align}
    \delta(\mathbf k,\theta)
    &=
    \left|
        \cos\!\left[
            \theta-\vartheta(\mathbf k)
        \right]
    \right|.
    \label{eq:app-rashba-delta-angle}
\end{align}
Thus each momentum favours the angle
\(\theta=\vartheta(\mathbf k)\), for which the magnitude of the
eigenvalues of \(\Gamma\) is maximal.

It remains to determine the range of \(\vartheta(\mathbf k)\).
Using
\begin{align}
    \tan\vartheta(\mathbf k)
    &=
    \frac{2ab}
    {E_+E_-+a^2+d^2-b^2},
    \label{eq:app-rashba-tan-vartheta}
\end{align}
together with
\begin{align}
    (E_+E_-)^2
    -
    \left(a^2+b^2-d^2\right)^2
    &=
    4a^2d^2
    \geq 0,
\end{align}
gives
\begin{align}
    E_+E_-+a^2+d^2-b^2
    &\geq
    2a^2.
\end{align}
For \(A>0\) and \(\lambda_{\rm R}\geq0\), Eq.~\eqref{eq:app-rashba-ratio}
therefore implies
\begin{align}
    0
    \leq
    \vartheta(\mathbf k)
    \leq
    \arctan\frac{\lambda_{\rm R}}{A}.
    \label{eq:app-rashba-vartheta-range}
\end{align}
The lower endpoint occurs at the time-reversal-invariant momenta,
where the spin-dependent terms vanish.
The upper endpoint occurs at momenta satisfying \(d(\mathbf k)=0\);
for the parameters used in the main text both endpoints are present
in the Brillouin zone.

The gap of the full operator \(\Gamma\) is the smallest value of
Eq.~\eqref{eq:app-rashba-delta-angle} over momentum.
For
\(0\leq\theta\leq\arctan(\lambda_{\rm R}/A)\), the two endpoints in
Eq.~\eqref{eq:app-rashba-vartheta-range} therefore give
\begin{align}
    \Delta_\Gamma(\theta)
    &=
    \min
    \left\{
        \cos\theta,\,
        \cos\!\left[
            \arctan\frac{\lambda_{\rm R}}{A}-\theta
        \right]
    \right\}.
    \label{eq:app-rashba-gap}
\end{align}
The first term decreases with \(\theta\), whereas the second increases.
The minimum is consequently maximal when the two endpoint gaps are
equal, which gives
\begin{align}
    \theta_{\rm opt}^{\rm R}
    &=
    \frac{1}{2}
    \arctan\frac{\lambda_{\rm R}}{A}.
    \label{eq:app-rashba-opt}
\end{align}
At this angle,
\begin{align}
    \Delta_\Gamma
    \bigl(\theta_{\rm opt}^{\rm R}\bigr)
    &=
    \cos\!\left[
        \frac{1}{2}
        \arctan\frac{\lambda_{\rm R}}{A}
    \right],
\end{align}
so the optimized auxiliary gap remains finite as the Rashba coupling
is increased.

\section{Analytic determination of \(S\) with Rashba--Dresselhaus spin-orbit coupling}
\label{app:Dress}

The translation-invariant square-lattice Hamiltonian with both Rashba
and Dresselhaus spin-orbit coupling is
\begin{align}
    H(\mathbf k)
    &=
    d(\mathbf k)\tau_z
    +
    2A
    \left(
        \sin k_x\,\tau_xs_x
        +
        \sin k_y\,\tau_xs_y
    \right)
    \nonumber\\
    &\quad
    +
    2\lambda_{\rm R}
    \left(
        \sin k_y\,s_x
        -
        \sin k_x\,s_y
    \right)
    \nonumber\\
    &\quad
    +
    2\lambda_{\rm D}
    \left(
        \sin k_x\,s_x
        -
        \sin k_y\,s_y
    \right),
    \label{eq:app-rd-H}
\end{align}
where
\begin{align}
    d(\mathbf k)
    &=
    M+2A(\cos k_x+\cos k_y).
    \label{eq:app-rd-d}
\end{align}
The matrices \(\tau_i\) act on the orbital degrees of freedom and
\(s_i\) on physical spin.
The term proportional to \(A\) is the original BHZ spin-orbit
coupling, while \(\lambda_{\rm R}\) and \(\lambda_{\rm D}\) control
the Rashba and Dresselhaus couplings, respectively.
Both spin-mixing terms preserve time-reversal symmetry.

It is convenient to introduce
\begin{align}
    p_x &= \sin k_x,
    &
    p_y &= \sin k_y,
\end{align}
and write Eq.~\eqref{eq:app-rd-H} as
\begin{align}
    H(\mathbf k)
    &=
    d\,\tau_z
    +
    \tau_x\,\mathbf g\cdot\mathbf s
    +
    \mathbf h\cdot\mathbf s ,
    \label{eq:app-rd-vector-H}
\end{align}
with
\begin{align}
    \mathbf g
    &=
    2A
    \begin{pmatrix}
        p_x\\
        p_y
    \end{pmatrix},
    \\
    \mathbf h
    &=
    2
    \begin{pmatrix}
        \lambda_{\rm R}p_y+\lambda_{\rm D}p_x\\
        -\lambda_{\rm R}p_x-\lambda_{\rm D}p_y
    \end{pmatrix}.
    \label{eq:app-rd-gh}
\end{align}
For pure Rashba coupling the two in-plane spin textures are
orthogonal at every momentum.
Dresselhaus coupling removes this property, since
\begin{align}
    \mathbf g\cdot\mathbf h
    &=
    4A\lambda_{\rm D}
    \left(
        p_x^2-p_y^2
    \right).
    \label{eq:app-rd-gh-dot}
\end{align}
This quantity is nonzero at generic momentum when
\(\lambda_{\rm D}\neq0\).
The simple conserved sector operator available for pure Rashba
coupling therefore no longer applies.

\subsection{Local grading operators}

We first determine whether Dresselhaus coupling requires a more general
local grading operator.
The onsite Hilbert space contains two orbital and two spin degrees of
freedom, so a Hermitian onsite operator can be expanded in the sixteen
matrices \(\tau_\mu s_\nu\), with
\(\mu,\nu=0,x,y,z\).
Time reversal acts as
\begin{align}
    T
    &=
    \i s_y K ,
\end{align}
and the grading must satisfy
\begin{align}
    TST^{-1}
    &=
    -S.
\end{align}
The ten Hermitian matrices that are odd under time reversal are
\begin{align}
    s_i,\qquad
    \tau_xs_i,\qquad
    \tau_zs_i,\qquad
    \tau_y,
    \qquad i=x,y,z.
\end{align}

The Rashba analysis identifies the local family
\begin{align}
    S(\theta)
    &=
    \cos\theta\,\tau_zs_z
    +
    \sin\theta\,\tau_y .
    \label{eq:app-rd-S-theta}
\end{align}
Both matrices in Eq.~\eqref{eq:app-rd-S-theta} square to unity and
anticommute, so \(S(\theta)^2=\id\) for every \(\theta\).
To determine whether Dresselhaus coupling introduces additional local
directions, consider an infinitesimal deformation
\(S\rightarrow S+\delta S\).
Preserving the involution condition to linear order requires
\begin{align}
    \{S,\delta S\}
    &=
    0.
\end{align}
Within the ten time-reversal-odd matrices, this condition leaves six
independent tangent directions,
\begin{align}
    G_1
    &=
    \sin\theta\,\tau_xs_x
    +
    \cos\theta\,s_y,
    \\
    G_2
    &=
    \sin\theta\,\tau_xs_y
    -
    \cos\theta\,s_x,
    \\
    G_3
    &=
    \tau_xs_z,
    \\
    G_4
    &=
    \tau_zs_x,
    \\
    G_5
    &=
    \tau_zs_y,
    \\
    G_6
    &=
    \cos\theta\,\tau_y
    -
    \sin\theta\,\tau_zs_z.
    \label{eq:app-rd-tangent}
\end{align}
The final direction is simply
\begin{align}
    G_6
    &=
    \partial_\theta S(\theta),
\end{align}
and therefore only changes the angle already present in
Eq.~\eqref{eq:app-rd-S-theta}.

The square-lattice Rashba-Dresselhaus Hamiltonian also preserves the
twofold rotation \(C_{2z}\).
Under this symmetry, \(G_1,G_2,G_4,\) and \(G_5\) are odd, whereas
\(G_3\) and \(G_6\) are even.
Restricting the grading to the same symmetry therefore leaves
\(G_3=\tau_xs_z\) as the only additional local direction beyond
changing \(\theta\).
The corresponding normalized two-angle family is
\begin{align}
    S(\theta,\phi)
    &=
    \cos\phi\,S(\theta)
    +
    \sin\phi\,\tau_xs_z.
    \label{eq:app-rd-S-two-angle}
\end{align}
The three matrices
\(\tau_zs_z\), \(\tau_y\), and \(\tau_xs_z\) square to unity and
mutually anticommute, so
\(S(\theta,\phi)^2=\id\).

The additional direction cannot increase the spectral gap of
\(\Gamma\).
For
\begin{align}
    G_3
    &=
    \tau_xs_z,
\end{align}
every term in Eq.~\eqref{eq:app-rd-vector-H} anticommutes with \(G_3\),
and hence
\begin{align}
    \{H,G_3\}
    &=
    0.
    \label{eq:app-rd-G3-anticommute}
\end{align}
At half filling, Eq.~\eqref{eq:app-rd-G3-anticommute} exchanges
occupied and unoccupied states and gives
\begin{align}
    G_3\rho G_3
    &=
    \id-\rho ,
    \\
    \{\rho,G_3\}
    &=
    G_3.
    \label{eq:app-rd-rho-G3}
\end{align}
Using
\(\Gamma=\{\rho,S\}-S\), Eq.~\eqref{eq:app-rd-S-two-angle} therefore
gives
\begin{align}
    \Gamma(\theta,\phi)
    &=
    \cos\phi\,\Gamma(\theta).
\end{align}
Consequently,
\begin{align}
    \Delta_\Gamma(\theta,\phi)
    &=
    |\cos\phi|\,
    \Delta_\Gamma(\theta),
\end{align}
which is maximal at \(\phi=0\).
The optimal grading therefore remains within the one-parameter family
\(S(\theta)\) in Eq.~\eqref{eq:app-rd-S-theta}.

\subsection{Auxiliary spectrum}

We next determine the optimal value of \(\theta\).
At half filling, provided the physical Hamiltonian remains gapped,
define the flattened Hamiltonian
\begin{align}
    Q(\mathbf k)
    &=
    \id-2\rho(\mathbf k)
    =
    \operatorname{sgn}H(\mathbf k).
\end{align}
The auxiliary operator can then be written as
\begin{align}
    \Gamma(\mathbf k,\theta)
    &=
    -\frac{1}{2}
    \left\{
        Q(\mathbf k),S(\theta)
    \right\}.
    \label{eq:app-rd-Gamma-Q}
\end{align}

Although Dresselhaus coupling removes the simple Rashba sector
operator, the Hamiltonian still admits an exact sector decomposition.
Squaring Eq.~\eqref{eq:app-rd-vector-H} gives
\begin{align}
    H^2
    &=
    \alpha\,\id
    +
    2B,
    \label{eq:app-rd-H2}
    \\
    \alpha
    &=
    d^2+|\mathbf g|^2+|\mathbf h|^2,
    \\
    B
    &=
    d\,\tau_z\,\mathbf h\cdot\mathbf s
    +
    (\mathbf g\cdot\mathbf h)\tau_x.
\end{align}
The two terms in \(B\) anticommute, giving
\begin{align}
    B^2
    &=
    R^2\id,
    \\
    R^2
    &=
    d^2|\mathbf h|^2
    +
    (\mathbf g\cdot\mathbf h)^2.
    \label{eq:app-rd-R}
\end{align}
One also finds \([B,H]=0\).
For \(R\neq0\), the normalized operator
\begin{align}
    K
    &=
    \frac{B}{R}
\end{align}
therefore satisfies \(K^2=\id\) and \([K,H]=0\).
The corresponding projectors are
\begin{align}
    \Pi_\pm
    &=
    \frac{\id\pm K}{2},
\end{align}
and the two positive energy scales are
\begin{align}
    E_\pm
    &=
    \sqrt{\alpha\pm2R}.
    \label{eq:app-rd-Epm}
\end{align}
The flattened Hamiltonian consequently takes the form
\begin{align}
    Q
    &=
    H
    \left(
        \frac{\Pi_+}{E_+}
        +
        \frac{\Pi_-}{E_-}
    \right).
    \label{eq:app-rd-Q}
\end{align}
The apparent singularity of \(K=B/R\) at \(R=0\) does not produce a
singularity in \(Q\): the two sectors become degenerate there, and
Eq.~\eqref{eq:app-rd-Q} has a smooth limit as long as the physical
energy gap remains open.

To evaluate Eq.~\eqref{eq:app-rd-Gamma-Q}, write
\begin{align}
    \frac{\Pi_+}{E_+}
    +
    \frac{\Pi_-}{E_-}
    &=
    u\,\id
    +
    v B,
\end{align}
where
\begin{align}
    u
    &=
    \frac{1}{2}
    \left(
        \frac{1}{E_+}
        +
        \frac{1}{E_-}
    \right),
    \\
    v
    &=
    \frac{1}{2R}
    \left(
        \frac{1}{E_+}
        -
        \frac{1}{E_-}
    \right).
\end{align}
Multiplication by \(H\) then gives
\begin{align}
    Q
    &=
    q_0\tau_z
    +
    \tau_x\,\mathbf a\cdot\mathbf s
    +
    \mathbf b\cdot\mathbf s
    +
    q_3\tau_ys_z,
    \label{eq:app-rd-Q-compact}
\end{align}
with
\begin{align}
    q_0
    &=
    d\left(u+v|\mathbf h|^2\right),
    \\
    \mathbf a
    &=
    u\mathbf g
    +
    v(\mathbf g\cdot\mathbf h)\mathbf h,
    \\
    \mathbf b
    &=
    u\mathbf h
    +
    v\left[
        d^2\mathbf h
        +
        (\mathbf g\cdot\mathbf h)\mathbf g
    \right],
    \\
    q_3
    &=
    vd(\mathbf g\times\mathbf h)_z.
\end{align}

Substituting Eq.~\eqref{eq:app-rd-Q-compact} into
Eq.~\eqref{eq:app-rd-Gamma-Q} gives
\begin{align}
    \Gamma(\mathbf k,\theta)
    &=
    -
    \left[
        \mathcal A_\theta s_z
        +
        \tau_y\,\mathbf C_\theta\cdot\mathbf s
    \right],
\end{align}
where
\begin{align}
    \mathcal A_\theta
    &=
    q_0\cos\theta
    +
    q_3\sin\theta,
    \\
    \mathbf C_\theta
    &=
    \cos\theta
    \begin{pmatrix}
        a_y\\
        -a_x
    \end{pmatrix}
    +
    \sin\theta
    \begin{pmatrix}
        b_x\\
        b_y
    \end{pmatrix}.
\end{align}
Since \(s_z\) anticommutes with the in-plane spin matrices,
\(\Gamma^2\) is proportional to the identity,
\begin{align}
    \Gamma^2(\mathbf k,\theta)
    &=
    \delta^2(\mathbf k,\theta)\,\id.
\end{align}
Using \(Q^2=\id\), the angular dependence reduces to
\begin{align}
    \delta^2(\mathbf k,\theta)
    &=
    \frac{1}{2}
    \left[
        1
        +
        \cos 2\vartheta(\mathbf k)\cos2\theta
        +
        \sin 2\vartheta(\mathbf k)\sin2\theta
    \right],
    \label{eq:app-rd-delta2}
\end{align}
where
\begin{align}
    \cos2\vartheta(\mathbf k)
    &=
    \frac{
        d^2+|\mathbf g|^2-|\mathbf h|^2
    }{
        E_+E_-
    },
    \\
    \sin2\vartheta(\mathbf k)
    &=
    \frac{
        -2(\mathbf g\times\mathbf h)_z
    }{
        E_+E_-
    }.
    \label{eq:app-rd-vartheta}
\end{align}
The normalization follows from
\begin{align}
    (E_+E_-)^2
    &=
    \left(
        d^2+|\mathbf g|^2-|\mathbf h|^2
    \right)^2
    +
    4(\mathbf g\times\mathbf h)_z^2.
\end{align}
Equation~\eqref{eq:app-rd-delta2} therefore becomes
\begin{align}
    \delta(\mathbf k,\theta)
    &=
    \left|
        \cos\!\left[
            \theta-\vartheta(\mathbf k)
        \right]
    \right|.
    \label{eq:app-rd-delta}
\end{align}
The eigenvalues of \(\Gamma(\mathbf k,\theta)\) are thus
\(-\delta,-\delta,+\delta,+\delta\).
For each momentum, their magnitude is maximal when
\(\theta=\vartheta(\mathbf k)\), which defines the preferred angle
at that momentum.

\subsection{Range of preferred angles}

It remains to determine the range of \(\vartheta(\mathbf k)\) across
the Brillouin zone.
For
\begin{align}
    r^2
    &=
    p_x^2+p_y^2
    \neq0,
\end{align}
define
\begin{align}
    \xi
    &=
    \frac{2p_xp_y}{p_x^2+p_y^2},
    &
    -1
    &\leq
    \xi
    \leq
    1,
    \\
    t
    &=
    \frac{d^2}{4(p_x^2+p_y^2)}
    \geq0.
\end{align}
The quantities entering Eq.~\eqref{eq:app-rd-vartheta} then become
\begin{align}
    |\mathbf g|^2
    &=
    4A^2r^2,
    \\
    |\mathbf h|^2
    &=
    4r^2
    \left(
        \lambda_{\rm R}^2
        +
        \lambda_{\rm D}^2
        +
        2\lambda_{\rm R}\lambda_{\rm D}\xi
    \right),
    \\
    (\mathbf g\times\mathbf h)_z
    &=
    -4Ar^2
    \left(
        \lambda_{\rm R}
        +
        \lambda_{\rm D}\xi
    \right).
    \label{eq:app-rd-invariants}
\end{align}
The angle \(2\vartheta\) is therefore the argument of the complex
number
\begin{align}
    Z(\xi,t)
    &=
    X(\xi,t)
    +
    \i Y(\xi),
\end{align}
with
\begin{align}
    X(\xi,t)
    &=
    t
    +
    A^2
    -
    \lambda_{\rm R}^2
    -
    \lambda_{\rm D}^2
    -
    2\lambda_{\rm R}\lambda_{\rm D}\xi,
    \\
    Y(\xi)
    &=
    2A
    \left(
        \lambda_{\rm R}
        +
        \lambda_{\rm D}\xi
    \right).
    \label{eq:app-rd-XY}
\end{align}

We restrict to
\begin{align}
    \lambda_{\rm D}^2
    &<
    A^2+\lambda_{\rm R}^2,
    \label{eq:app-rd-gap-condition}
\end{align}
together with parameters for which the physical band gap remains
open.
The upper endpoint follows by defining
\begin{align}
    \vartheta_+
    &=
    \arctan
    \frac{
        \lambda_{\rm R}+\lambda_{\rm D}
    }{A}.
\end{align}
Rotating \(Z\) by \(2\vartheta_+\) gives
\begin{align}
    \operatorname{Im}
    \left[
        Ze^{-2\i\vartheta_+}
    \right]
    &=
    \frac{
        2A
    }{
        A^2+(\lambda_{\rm R}+\lambda_{\rm D})^2
    }
    \nonumber\\
    &\quad\times
    \left[
        \lambda_{\rm D}(\xi-1)
        \left(
            A^2+\lambda_{\rm R}^2-\lambda_{\rm D}^2
        \right)
    \right.
    \nonumber\\
    &\hspace{1.4cm}\left.
        -
        (\lambda_{\rm R}+\lambda_{\rm D})t
    \right].
\end{align}
The right-hand side is non-positive because
\(\xi\leq1\), \(t\geq0\), and
Eq.~\eqref{eq:app-rd-gap-condition} holds.
Hence
\begin{align}
    \vartheta(\mathbf k)
    &\leq
    \arctan
    \frac{
        \lambda_{\rm R}+\lambda_{\rm D}
    }{A}.
    \label{eq:app-rd-vartheta-max}
\end{align}
The bound is reached for \(\xi=1\) and \(t=0\), corresponding to
\(p_x=p_y\) and \(d(\mathbf k)=0\).

The lower endpoint depends on the relative strengths of the two
spin-orbit couplings.
For
\(0\leq\lambda_{\rm D}\leq\lambda_{\rm R}\),
Eq.~\eqref{eq:app-rd-XY} gives \(Y(\xi)\geq0\) for all
\(\xi\), and therefore
\begin{align}
    \vartheta(\mathbf k)
    &\geq0.
\end{align}
The lower bound is attained at the time-reversal-invariant momenta,
where the spin-dependent terms vanish.

For
\(\lambda_{\rm D}>\lambda_{\rm R}\), define instead
\begin{align}
    \vartheta_-
    &=
    \arctan
    \frac{
        \lambda_{\rm R}-\lambda_{\rm D}
    }{A}
    <0.
\end{align}
A rotation by \(2\vartheta_-\) gives
\begin{align}
    \operatorname{Im}
    \left[
        Ze^{-2\i\vartheta_-}
    \right]
    &=
    \frac{
        2A
    }{
        A^2+(\lambda_{\rm R}-\lambda_{\rm D})^2
    }
    \nonumber\\
    &\quad\times
    \left[
        \lambda_{\rm D}(\xi+1)
        \left(
            A^2+\lambda_{\rm R}^2-\lambda_{\rm D}^2
        \right)
    \right.
    \nonumber\\
    &\hspace{1.4cm}\left.
        +
        (\lambda_{\rm D}-\lambda_{\rm R})t
    \right],
\end{align}
which is non-negative under the same conditions.
Hence
\begin{align}
    \vartheta(\mathbf k)
    &\geq
    \arctan
    \frac{
        \lambda_{\rm R}-\lambda_{\rm D}
    }{A}.
    \label{eq:app-rd-vartheta-min-negative}
\end{align}
This bound is reached for \(\xi=-1\) and \(t=0\), corresponding to
\(p_y=-p_x\) and \(d(\mathbf k)=0\).

Provided the contour \(d(\mathbf k)=0\) intersects both momentum-space
diagonals, the endpoint values are therefore
\begin{align}
    \vartheta_{\rm max}
    &=
    \arctan
    \frac{
        \lambda_{\rm R}+\lambda_{\rm D}
    }{A},
    \\
    \vartheta_{\rm min}
    &=
    \begin{cases}
        0,
        &
        0\leq\lambda_{\rm D}\leq\lambda_{\rm R},
        \\[2mm]
        \displaystyle
        \arctan
        \frac{
            \lambda_{\rm R}-\lambda_{\rm D}
        }{A},
        &
        \lambda_{\rm R}<\lambda_{\rm D}
        <
        \sqrt{A^2+\lambda_{\rm R}^2}.
    \end{cases}
    \label{eq:app-rd-vartheta-range}
\end{align}
For
\(d(\mathbf k)=M+2A(\cos k_x+\cos k_y)\),
the contour \(d(\mathbf k)=0\) intersects both diagonals when
\(|M|<4A\).
For \(M=-2A\), for example, the intersections occur at
\(k_x=\pm k_y=\pm\pi/3\).

The condition in Eq.~\eqref{eq:app-rd-gap-condition} also has a
physical interpretation.
At
\(\lambda_{\rm D}^2=A^2+\lambda_{\rm R}^2\),
the two in-plane spin textures can become parallel with equal
magnitude on the contour \(d(\mathbf k)=0\), allowing the lower
physical energy \(E_-\) to vanish.

\subsection{Optimal angle}

For a fixed onsite operator \(S(\theta)\), the spectral gap of
\(\Gamma\) is
\begin{align}
    \Delta_\Gamma(\theta)
    &=
    \min_{\mathbf k}
    \left|
        \cos[
            \theta-\vartheta(\mathbf k)
        ]
    \right|.
\end{align}
All preferred angles lie in
\([\vartheta_{\rm min},\vartheta_{\rm max}]\), and both endpoints are
attained under the conditions stated above.
The smallest gap is therefore set by the endpoint furthest from the
chosen value of \(\theta\).
The maximum of this minimum occurs when the distances to the two
endpoints are equal,
\begin{align}
    \theta-\vartheta_{\rm min}
    &=
    \vartheta_{\rm max}-\theta.
\end{align}
The optimal angle is consequently
\begin{align}
    \theta_{\rm opt}^{\rm D}
    &=
    \frac{
        \vartheta_{\rm min}
        +
        \vartheta_{\rm max}
    }{2}
    \nonumber\\
    &=
    \frac{1}{2}
    \left[
        \arctan
        \frac{
            \lambda_{\rm R}+\lambda_{\rm D}
        }{A}
    \right.
    \nonumber\\
    &\hspace{1.3cm}\left.
        +
        \arctan
        \frac{
            \min(0,\lambda_{\rm R}-\lambda_{\rm D})
        }{A}
    \right].
    \label{eq:app-rd-theta-opt}
\end{align}
The corresponding auxiliary gap is
\begin{align}
    \Delta_\Gamma^{\rm opt}
    &=
    \cos
    \left[
        \frac{
            \vartheta_{\rm max}
            -
            \vartheta_{\rm min}
        }{2}
    \right].
\end{align}
The optimized local grading is therefore
\begin{align}
    S_{\rm opt}
    &=
    \cos\theta_{\rm opt}^{\rm D}\,
    \tau_zs_z
    +
    \sin\theta_{\rm opt}^{\rm D}\,
    \tau_y.
\end{align}
Setting \(\lambda_{\rm D}=0\) recovers the pure Rashba result
\begin{align}
    \theta_{\rm opt}^{\rm R}
    &=
    \frac{1}{2}
    \arctan
    \frac{\lambda_{\rm R}}{A}.
\end{align}

\section{Finding an $S$ through numeric optimization}
\label{app:S-optimization}

When $\rho$ is less strongly localized, a random choice of the unit vectors in Eq.~\eqref{eq:S-local-kramers-ansatz} need no longer keep the auxiliary spectrum sufficiently gapped. 
We therefore need to optimize the unit vectors such that $\Gamma^2$ is as close to the identity as possible, avoiding eigenvalues near zero.
We use the cost function\footnote{The log-determinant cost is smooth and basis independent on the gapped domain. 
It diverges whenever an eigenvalue of $\Gamma$ approaches zero and admits a convex, self-concordant extension (meaning that its third derivatives are controlled by its Hessian, which underpins convergence guarantees for Newton methods) to unconstrained Hermitian $S$ on $\|[S,\rho]\|_{\mathrm{op}}<1$, although the constraint $S^2=1$ makes the optimization nonconvex; see for instance Ref.~\cite{boyd2004}.}
\begin{multline}
    \mathcal F(S)
    =-\log\det\Gamma^2\\
      =-\operatorname{Tr}\log\bigl(1+[S,\rho]^2\bigr)
      =-\sum_j\log\lambda_j^2,
    \label{eq:S-log-cost}
\end{multline}
regarded as a functional of the local operator $S$, where $\lambda_j$ are the eigenvalues of $\Gamma$. 
Eq.~\eqref{eq:S-log-cost} is a smooth, basis-independent function on the gapped domain that penalizes each eigenvalue individually. 
Since $\lambda_j^2\in[0,1]$, the cost vanishes precisely when all squared eigenvalues are unity while diverging as any eigenvalue approaches zero. 
It therefore simultaneously favors $\tilde S^2$ being the identity, and strongly suppresses nearly singular solutions.

As a starting point, we use the truncated cost function,
\begin{align}
    \mathcal F(S)
    &=\underbrace{\|[S,\rho]\|_F^2}_{\mathcal F_2(S)}
      +\mathcal O\bigl(\|[S,\rho]\|_F^4\bigr),
    \label{eq:S-truncated-cost}
\end{align}
where $\|X\|_F=\sqrt{\operatorname{Tr}(X^\dagger X)}$ denotes the Frobenius (Hilbert--Schmidt) norm.
Writing $\Sigma_{K\alpha}^{\mathbf r}=W_{\mathbf rK}\sigma_\alpha W_{\mathbf rK}^{\dagger}$ and collecting the unit vectors into $N=\bigoplus_{\mathbf r,K}\hat{\mathbf n}_{\mathbf rK}$, the direct terms in the squared commutator give a constant with respect to $N$. 
Since the onsite commutators vanish, summing the off-diagonal contributions gives
\begin{align}
    \mathcal F_2&=\mathcal F_0+N^TJN,
    \qquad
    \mathcal F_0=2\sum_{\mathbf r\ne\mathbf r'}
                     \|\rho_{\mathbf r\mathbf r'}\|_F^2,
    \label{eq:S-classical-spin-cost}\\
    J_{(\mathbf rK\alpha),(\mathbf r'K'\beta)}
    &=-2\operatorname{Re}\operatorname{Tr}\bigl(
       \Sigma_{K\alpha}^{\mathbf r}\rho_{\mathbf r\mathbf r'}
       \Sigma_{K'\beta}^{\mathbf r'}\rho_{\mathbf r'\mathbf r}
       \bigr),\qquad \mathbf r\ne\mathbf r',
\end{align}
with the blocks of $J$ at $\mathbf r=\mathbf r'$ set to zero.
Thus minimizing the truncated cost amounts to finding the ground state of a classical spin model, with one three-component unit vector for each local Kramers pair. 
In a disordered system the couplings can be anisotropic and frustrated, giving a classical spin-glass problem. 
Moreover, since $\|\rho_{\mathbf r\mathbf r'}\|\leq C e^{-|\mathbf r-\mathbf r'|/\xi}$, the couplings decay as $|J_{(\mathbf rK\alpha),(\mathbf r'K'\beta)}|\leq\tilde C e^{-2|\mathbf r-\mathbf r'|/\xi}$.
The resulting classical model therefore has local interactions.

A simple mean-field starting point is obtained by relaxing the individual normalization conditions to a single global constraint,
\begin{equation}
    \sum_{\mathbf r,K}|\mathbf n_{\mathbf rK}|^2=M,
    \qquad M=\frac12\sum_{\mathbf r}d_{\mathbf r}.
\end{equation}
The relaxed minimum is given by an eigenvector corresponding to the smallest eigenvalue of $J$, normalized to length $\sqrt M$.
If several eigenvectors have approximately the same eigenvalue, we can choose a linear combination with the most uniform local norms. 
We normalize each vector individually and improve the configuration by sweeping over the spins, minimizing with respect to one vector at a time while keeping the others fixed. 
With the above convention for $J$, each such update simply aligns the spin opposite to its local field,
\begin{equation}
    \hat{\mathbf n}_a\longleftarrow
    -\frac{\mathbf h_a}{|\mathbf h_a|},
    \qquad
    \mathbf h_a=\sum_{b\ne a}J_{ab}\hat{\mathbf n}_b,
    \qquad a=(\mathbf r,K),
\end{equation}
where $J_{ab}$ denotes a $3\times3$ block of $J$ and the update is performed whenever $\mathbf h_a\ne0$. 
This provides an inexpensive approximate solution to the truncated optimization problem.
The difference between the achieved cost (after renormalization) and the relaxed lower bound $\mathcal F_0+M\lambda_{\min}(J)$ bounds the error in the minimum cost from above, providing an explicit check of the accuracy for each instance.

Even an accurate minimization of the truncated cost is not necessarily sufficient.
Since $\mathcal F_2=\sum_j(1-\lambda_j^2)$, it favors increasing all squared eigenvalues with equal weight and does not disfavor eigenvalues close to zero strongly enough.
It can therefore leave a small gap, or even a gapless auxiliary spectrum.
In that case the spin-glass optimization can still be used as a starting point and then refine it using the full logarithmic cost.

The remaining optimization is close to a convex problem in a precise sense, since
\begin{equation}
    \mathcal F(S)
    =-\log\det(1-\i[S,\rho])-\log\det(1+\i[S,\rho]).
    \label{eq:S-convex-extension}
\end{equation}
On the convex domain where the eigenvalues of $\i[S,\rho]$ lie strictly between $-1$ and $1$, the right-hand side is convex and self-concordant (meaning that its third derivatives are controlled by its Hessian, making it particularly well suited to Newton's method) as a function of the unconstrained Hermitian matrix $S$, since $\i[S,\rho]$ depends linearly on $S$.
It agrees with the full cost on $S^2=1$ and defines a convex extension away from this constraint.
In the parametrization \eqref{eq:S-local-kramers-ansatz}, the only nonconvex constraints are the unit-vector normalizations.

To optimize efficiently, we use the gradient and Hessian of this extension with respect to the Hilbert--Schmidt inner product.
Writing $C=[S,\rho]$, $R=(1+C^2)^{-1}=\tilde S^{-2}$ on the ansatz, and $\mathcal C_O(X)=OXO$, we obtain
\begin{align}
    \nabla_S\mathcal F
    &=-2[\rho,CR],
    \label{eq:S-log-gradient}\\
    \mathcal H_S^{\mathcal F}(X)
    &=-2i\left[\rho,
       \left(\mathcal C_R+\mathcal C_{iCR}\right)
       \bigl(i[\rho,X]\bigr)\right].
    \label{eq:S-log-hessian}
\end{align}
The Hessian is symmetric and positive semidefinite with respect to this inner product, consistently with \eqref{eq:S-convex-extension}. 
To impose the normalization constraints, we use local coordinates $x$ for the unit vectors, for example by normalizing $\hat{\mathbf n}_a+\delta\mathbf n_a$
with $\delta\mathbf n_a\perp\hat{\mathbf n}_a$. 
With $E_i=\partial_iS$ and $E_{ij}=\partial_i\partial_jS$, the gradient and Hessian in these coordinates are
\begin{align}
    g_i&=\operatorname{Tr}\bigl(E_i\nabla_S\mathcal F\bigr),\\
    H_{ij}
    &=\operatorname{Tr}\bigl(E_i\mathcal H_S^{\mathcal F}(E_j)\bigr)
      +\operatorname{Tr}\bigl(E_{ij}\nabla_S\mathcal F\bigr).
    \label{eq:S-constrained-hessian}
\end{align}
The second term accounts for the curvature introduced by the normalization constraints and is the reason for why the constrained problem need not remain convex.
The purpose of the preceding spin-glass optimization is to bring the initial configuration close to a minimum of the full cost function.
When the higher-order corrections to the gradient and Hessian are sufficiently small near a nondegenerate minimum of the truncated cost, the full cost function has a nearby local minimum. 
We expect the initial configuration to lie in a neighborhood containing this minimum in which the constrained Hessian is positive definite.

When the initialization lies sufficiently close to a minimum of the full cost function, the constrained Hessian is positive definite and Newton's method converges quadratically. 
We solve
\begin{equation}
    Hv=-g
\end{equation}
and perform a line search along the corresponding normalized update.
If the constrained Hessian is not positive definite, a regularized Newton step or a gradient step can be used until this local regime is reached.
Starting from a good spin-glass configuration, only a few such refinement steps may therefore be needed to approach a local optimum of the full cost within the chosen parametrization.

\bibliography{refs}

\end{document}